\documentclass[11pt]{article}
\usepackage{epsfig}
\usepackage{mathtools}
\usepackage{amsfonts}
\usepackage{amssymb}
\usepackage{amsmath}
\usepackage{listings}

\usepackage{color}
\usepackage{slashed}
\usepackage{cite}
\usepackage{caption}
\usepackage{subcaption}

\usepackage{mmacells}

\usepackage{geometry} 
\usepackage{graphicx}
\graphicspath{{pictures/}}

\usepackage{multirow}           
\usepackage{array} 

\definecolor{mygray}{rgb}{0.5,0.5,0.5}
\definecolor{mymauve}{rgb}{0.58,0,0.82}

\allowdisplaybreaks

\def\ep  {\varepsilon}
\def\E4  {\mathcal{E}_4}

\renewcommand{\Re}{\mathop{\mathrm{Re}}\nolimits}
\renewcommand{\Im}{\mathop{\mathrm{Im}}\nolimits}

\newcommand*\pFqskip{6mu}
\catcode`,\active
\newcommand*\pFq{\begingroup
	\catcode`\,\active
	\def ,{\mskip\pFqskip\relax}%
	\dopFq
}
\catcode`\,12
\def\dopFq#1#2#3#4#5{%
	{}_{#1}F_{#2}\biggl[\genfrac..{0pt}{}{#3}{#4};#5\biggr]%
	\endgroup
}

\newcommand*\Fkdfskip{6mu}
\catcode`,\active
\newcommand*\Fkdf{\begingroup
	\catcode`\,\active
	\def ,{\mskip\Fkdfskip\relax}%
	\doFkdf
}
\catcode`\,12
\def\doFkdf#1#2#3#4#5#6#7#8#9{%
	F^{#1}_{#2}\biggl[\genfrac..{0pt}{}{#3}{#4} \Big| \genfrac..{0pt}{}{#5}{#6} \Big| \genfrac..{0pt}{}{#7}{#8};#9\biggr]%
	\endgroup
}

\begin{document}

\begin{center}

\vspace{3cm}

{\bf \Large Numerical analytical continuation of multivariate hypergeometric functions} \vspace{1cm}

{\large M.A. Bezuglov$^{1}$, B.A. Kniehl$^{1}$, A.I. Onishchenko$^{2}$ and O.L. Veretin$^{3}$}\vspace{0.5cm}

{\it $^1$ II.~Institut f\"ur Theoretische Physik, Universit\"at Hamburg, Hamburg, Germany \\	
$^2$ Bogoliubov Laboratory of Theoretical Physics, Joint
Institute for Nuclear Research, \\ Dubna, Russia, \\
$^3$ Institut f\"ur Theoretische Physik, Universit\"at Regensburg, Regensburg, Germany}
\vspace{1cm}
\end{center}

\begin{abstract}
	 We present a general framework for the high-precision numerical evaluation of multivariate hypergeometric functions defined as solutions of holonomic systems of partial differential equations. Our approach adapts and extends methods originally developed for multi-loop Feynman integrals to the setting of hypergeometric functions of many variables. In particular, we construct Pfaffian systems for arbitrary multivariate hypergeometric functions by applying the Laporta reduction algorithm to suitable systems of differential relations. Next, we construct a numerical scheme based on the Frobenius method, which allows us to compute local power-series solutions with controlled precision and to transport them along prescribed paths in the space of variables. A central part of the paper is devoted to a systematic analysis of multivaluedness and branch structure: we show how the Frobenius method can be used to access different Riemann sheets in a controlled way and to track changes of the solution under analytic continuation around singular loci.
\end{abstract}

\begin{center}
\end{center}

\newpage

\tableofcontents{}\vspace{0.5cm}

\renewcommand{\theequation}{\thesection.\arabic{equation}}

\section{Introduction}
\label{sec:intro}

Hypergeometric functions of several variables form a remarkably rich class of special functions with applications ranging from algebraic geometry and number theory to statistics and high-energy physics. Many multivariate hypergeometric functions including the Appell and Lauricella functions and their generalizations arise naturally as solutions of holonomic systems of linear partial differential equations. In multi-loop Feynman integral computations, such functions enter both explicitly, via closed-form representations, and implicitly, through the differential equations satisfied by the integrals \cite{Weinzierl:2022eaz,Dubovyk:2022obc,smirnov2006feynman, Tarasov:2006nk, Lee:2009dh, Lee:2012hp, Tarasov:2022clb, Bezuglov:2022npo, Bezuglov:2021tax, Blumlein:2021hbq}. This makes robust and flexible tools for their high-precision numerical evaluation highly desirable in both mathematics and physics. Early investigations of the interplay between loop integrals and hypergeometric functions include \cite{Kershaw:1973km,Kreimer:1991wj,Kreimer:1992ps,Brucher:1993sp,Boos:1990rg,Davydychev:1990cq,Broadhurst:1993mw}, and a broader overview of these connections is available in \cite{Kalmykov:2020cqz}.

In typical applications, the parameters of the relevant hypergeometric functions depend linearly on the dimensional-regularization parameter $\ep$, and one is therefore interested in their Laurent expansions in $\ep$ to a prescribed order. A variety of methods and public codes exist to obtain such expansions—both analytically and numerically for different classes of hypergeometric functions \cite{Huber:2005yg,Huber:2007dx,Moch:2005uc,Weinzierl:2002hv,Ablinger:2013cf,Huang:2012qz,Bera:2023pyz,Bezuglov:2023owj,Greynat:2013hox,Greynat:2014jsa,Borowka:2017idc}.

Analytic continuation beyond the domain of absolute convergence is a central practical issue. Broadly speaking, the literature follows two complementary routes. The first relies on connection formulas and rearrangements of multivariate series, often reducing the continuation problem to known continuation relations for functions of fewer variables (or to special subfamilies where transformation theory is well developed) \cite{AppelKampedeFeriet,Erdelyi,Olsson,Exton,Ananthanarayan:2024nsc,Bera:2024hlq}. The second route proceeds from integral representations; most notably Mellin--Barnes contours \cite{hahne1969analytic,Bezrodnykh1,Bezrodnykh2,Bezrodnykh3,Bezrodnykh4,Bezrodnykh5,BezrodnykhReview,HYPERDIRE1,HYPERDIRE2,HYPERDIRE3,HYPERDIRE4}.

From the point of view of differential equations, a multivariate hypergeometric function is characterized by a holonomic system with regular (or mildly irregular) singularities along an algebraic locus in the space of variables and parameters \cite{SaitoSturmfelsTakayama}. For practical purposes, one often prefers to work with an equivalent first-order system in Pfaffian form, which is particularly well suited for numerical propagation, analytic continuation and the study of monodromy. In the context of Feynman integrals, powerful techniques have been developed to derive and manipulate such systems, most notably the differential-equations \cite{diffeqn1,diffeqn2,diffeqn3,diffeqn4,diffeqn5,epform1,epform2} and integration-by-parts (IBP) methods \cite{IBP1,IBP2}. These techniques allow one to express derivatives of a given integral in terms of a finite basis of master integrals and thereby obtain a closed system of first-order differential equations for the basis.

In our previous work on Lauricella-type functions \cite{Bezuglov:2025xol}, we exploited precisely this connection between hypergeometric functions and holonomic systems. There we focused on specific Lauricella hypergeometric functions and constructed a numerical algorithm based on the Frobenius method. The key ingredients were: (i) the derivation of a suitable Pfaffian system for the chosen family of functions; (ii) the computation of local Frobenius solutions around regular singular points with arbitrary precision; and (iii) an efficient strategy for analytic continuation along prescribed paths in the space of variables. In a subsequent paper \cite{Bezuglov:2025msm}, we implemented this algorithm in a public computer program, providing a practical tool for the high-precision evaluation of Lauricella functions in several variables.

The aim of the present paper is to develop a general framework for the numerical evaluation of multivariate hypergeometric functions starting from their series representations. Given a hypergeometric series in several variables, we first derive the holonomic system of linear partial differential equations with rational coefficients that it satisfies, and then apply a Laporta-style reduction \cite{Laporta:2000dsw} to express all derivatives in terms of a finite basis, thereby obtaining a Pfaffian first-order system for the corresponding vector of basis functions. On this basis, we employ the Frobenius method to construct local series solutions around regular singular points with arbitrary precision and to use them as boundary data for numerical propagation along prescribed paths. Particular attention is paid to the analytic structure of the solutions: we show how this Frobenius-based approach allows one to systematically access different Riemann sheets, track the monodromy under continuation around components of the singular locus, and thereby perform consistent high-precision evaluations in multivalued settings.

The remainder of this paper is organized as follows. In Section~\ref{sec:hypergeom-holonomic} we introduce a general definition of multivariate hypergeometric series in terms of lattices and linear forms and recall the associated holonomic systems of partial differential equations. Section~\ref{sec:laporta} explains how to treat these systems in close analogy with IBP relations: we act with differential operators, set up a Laporta-style reduction for derivatives, identify a finite set of ``master derivatives'', and extract Pfaffian systems suitable for numerical integration. In Section~\ref{sec:frobenius} we describe the generalized Frobenius method used to construct local series solutions on a given Riemann sheet and glue them together along the prescribed path made from chain of overlapping discs avoiding cuts.
Section~\ref{sec:aplications} illustrates the practical use of this approach on a number of examples, in particular multi-loop Feynman integrals expressible in terms of multivariate hypergeometric functions. Section~\ref{sec:monodromies} then focuses on the monodromy structure of multivalued hypergeometric functions. After considering simple examples of hypergeometric functions of one and two variables we come with a systematic procedure for analytical continuation of multivariable hypergeometric functions across multiple Riemann sheets.
In Section~\ref{sec:confluent} we briefly discuss confluent limits and irregular singularities, outlining the extensions required to cover such cases. Finally, in Section~\ref{sec:conclusions} we present our conclusions. Appendix~\ref{appendix::HAPC} contains details on the usage of the \texttt{HAPC} package and its tests.

\label{key}

\section{Hypergeometric functions}
\label{sec:hypergeom-holonomic}

We begin by fixing the integer lattice
\[
L = \mathbb{Z}^n, \qquad 
e_1 = (1,0,\dots,0), \dots, e_n = (0,\dots,0,1),
\]
together with its dual lattice
\[
L^{\vee} = \text{Hom}_{\mathbb{Z}}(L,\mathbb{Z}),
\]
whose elements are $\mathbb{Z}$–linear functionals on $L$. For each index $k$ in two finite index sets $K_1$ and $K_2$ we choose linear forms
\[
s_k, r_k \in L^{\vee},
\]
and complex parameters $\alpha_k$ and $\beta_k$, respectively. The corresponding multivariate hypergeometric series is defined by \cite{Aomoto:2011ggg}
\begin{equation}
F\left( (\alpha_k);(\beta_k);x\right) = \sum\limits_{\nu} \frac{\prod\limits_{k \in K_1}(\alpha_k)_{s_k(\nu)}}{\prod\limits_{k \in K_2}(\beta_k)_{r_k(\nu)}}\frac{x^{\nu}}{\nu!}, \qquad s_k(\nu),r_k(\nu) \in L^{\vee},
\end{equation}
where $(\cdot)_m$ denotes the Pochhammer symbol and the sum runs over multi-indices
\begin{equation}
\nu = (\nu_1,\dots,\nu_n) \in \mathbb{Z}^n_{\ge 0}, \qquad  \nu! = \nu_1!\cdots \nu_n!, \qquad x = (x_1,\dots,x_n) \in \mathbb{C}^n, \qquad x^{\nu} = x_1^{\nu_1} \cdots x_n^{\nu_n}.
\end{equation}
Thus each coefficient of the series is a finite product of Pochhammer symbols whose arguments are integer–valued linear forms in $\nu$, divided by a similar product in the denominator, multiplied by the usual monomial $x^\nu/\nu!$.

The hypergeometric character of the series is encoded in a balancing condition on the linear forms $s_k$ and $r_k$. We require that for each basis vector $e_i$ one has
\begin{equation}
\label{eq:condition}
\sum\limits_{k\in K_1}s_k(e_i)-\sum\limits_{k\in K_2}r_k(e_i) = 1, \qquad 1\le i \le n.
\end{equation}
Equivalently, the ratio of neighbouring coefficients in the $i$–th direction,
\[
\frac{a_{\nu+e_i}}{a_{\nu}}, \qquad 
a_{\nu} = \frac{\prod\limits_{k \in K_1}(\alpha_k)_{s_k(\nu)}}{\prod\limits_{k \in K_2}(\beta_k)_{r_k(\nu)}}\frac{1}{\nu!},
\]
is a rational function of $\nu$ whose dependence on $\nu_i$ is of the hypergeometric type (linear in the shift). This condition guarantees, in particular, that the series coefficients satisfy a finite system of linear recurrence relations with polynomial coefficients, and that the resulting function is a solution of a holonomic system of linear partial differential equations with rational coefficients in $x$. In other words, $F\big((\alpha_k);(\beta_k);x\big)$ is a multivariate hypergeometric function in the sense relevant for this work.

The definition above is flexible enough to encompass many familiar examples. Suitable choices of the index sets $K_1$, $K_2$ and of the linear forms $s_k$, $r_k$ reproduce classical Horn-type functions, the Appell and Lauricella families, as well as various hypergeometric functions that appear in the representation of Feynman integrals. In this paper we adopt the following convention: by a “multivariate hypergeometric function’’ we mean a function represented locally by a convergent series of the form given above (for some choice of data $(K_1,K_2,s_k,r_k,\alpha_k,\beta_k)$), understood as a germ at the origin, together with its analytic continuation along paths in the domain where the associated holonomic system is defined. This provides the starting point for the differential–equation and Pfaffian framework developed in the subsequent sections.

Starting from this series representation, one can derive a system of linear partial differential equations with rational coefficients that is satisfied by $F\big((\alpha_k);(\beta_k);x\big)$. The construction is entirely algebraic and rests on the dependence of the coefficients on the multi-index $\nu$. Let
\[
a_\nu = \frac{\prod\limits_{k \in K_1}(\alpha_k)_{s_k(\nu)}}{\prod\limits_{k \in K_2}(\beta_k)_{r_k(\nu)}}\frac{1}{\nu!}
\]
denote the coefficient of $x^\nu$ in the series. By definition of the Pochhammer symbol and of the linear forms $s_k,r_k$, the ratio of neighbouring coefficients in the $i$–th direction can be written as a rational function of $\nu$,
\[
\frac{a_{\nu+e_i}}{a_\nu} = R_i(\nu)\,,
\]
where $R_i$ is an explicit rational expression whose numerator and denominator are products of linear forms in $\nu$. 

To turn these coefficient relations into differential equations it is convenient to introduce the Euler operators
\[
\theta_i = x_i \frac{\partial}{\partial x_i}\,, \qquad i=1,\dots,n\,,
\]
which act diagonally on monomials:
\[
\theta_i\,x^\nu = \nu_i\, x^\nu\,, \qquad
p(\theta_1,\dots,\theta_n)\,x^\nu = p(\nu_1,\dots,\nu_n)\,x^\nu
\]
for any polynomial $p$. Using the ratio $a_{\nu+e_i}/a_\nu$ and clearing denominators, we obtain polynomial relations of the form
\[
\sum_{j} p_{i,j}(\nu)\, a_{\nu+\delta_{i,j}} = 0\,,
\]
where the shifts $\delta_{i,j} \in \mathbb{Z}^n$ and the polynomials $p_{i,j}$ depend on the data $(K_1,K_2,s_k,r_k,\alpha_k,\beta_k)$. Multiplying this relation by $x^{\nu+\delta_{i,j}}$ and summing over all $\nu \in \mathbb{Z}_{\ge 0}^n$, we can replace each factor $p_{i,j}(\nu)$ by the corresponding polynomial $p_{i,j}(\theta)$ and each shift of the index by multiplication with a monomial in $x$. In this way we arrive at differential identities of the schematic form
\[
\mathcal{L}_\mu(x,\theta)\, F\big((\alpha_k);(\beta_k);x\big) = 0\,,
\]
where $\mathcal{L}_\mu$ are polynomials in the Euler operators with coefficients rational in $x$ and in the parameters. Different ways of combining the basic shifts and eliminating denominators produce a finite collection of such operators. 

By construction, these operators generate a holonomic system of partial differential equations: the space of local solutions has finite dimension, and $F\big((\alpha_k);(\beta_k);x\big)$ is one distinguished solution singled out by its series expansion at the origin. For concrete choices of the data $(K_1,K_2,s_k,r_k)$ one recovers the well-known differential equations of classical multivariate hypergeometric functions (for example, the Appell and Lauricella systems). In the present work we do not rely on any special structure of these examples; instead, we take the general holonomic system associated with the series as our starting point and apply the Laporta-style reduction and Frobenius-based numerical methods described in the following sections.

\section{Equations for hypergeometric functions}
\label{sec:laporta}

In this section we explain how the system of partial differential equations satisfied by a multivariate hypergeometric function can be treated in complete analogy with integration-by-parts (IBP) identities for Feynman integrals. The key idea is to view suitable combinations of derivatives of the hypergeometric function as analogues of individual Feynman integrals in a given family, and to use a Laporta-style reduction to express all such derivatives through a finite set of ``master derivatives''. This set will then form the basis of the vector of functions that obeys the Pfaffian system used for numerical evaluation.

\subsection{Laporta algorithm for Feynman integrals}

In the standard setting of multi-loop calculations, one considers a family of Feynman integrals
\begin{equation}
  I(\vec{a}) \equiv I(a_1,\dots,a_N)
  = \int \frac{\mathrm{d}^d k_1 \cdots \mathrm{d}^d k_L}{D_1^{a_1} \cdots D_N^{a_N}} \,,
  \label{eq:Laporta-family}
\end{equation}
where $D_j$ are propagator denominators and the $a_j \in \mathbb{Z}$ are their powers. Integration-by-parts identities and, where applicable, Lorentz-invariance identities yield linear relations of the form
\begin{equation}
  \sum_{\vec{a}} c_{\vec{a}}(d, \{p_i\}) \, I(\vec{a}) = 0 \,,
  \label{eq:IBP-relations}
\end{equation}
with coefficients $c_{\vec{a}}$ rational in the kinematic invariants and in the dimensional-regularization parameter $d$. The Laporta algorithm provides a systematic procedure to solve this overconstrained linear system and express all integrals $I(\vec{a})$ in the family in terms of a finite set of basis elements, the so-called master integrals.

The central ingredients of the Laporta \cite{Laporta:2000dsw, Maierhofer:2017gsa} method are \footnote{There are other packages that use the Laporta method, for example \cite{Maierhofer:2018gpa,Lange:2025fba,Anastasiou:2004vj,Guan:2024byi,Smirnov:2008iw, Smirnov:2013dia,Smirnov:2014hma,Studerus:2009ye,vonManteuffel:2012np}.}:

\begin{itemize}

  \item A \emph{ranking} step. Each integral is labelled by the multi-index $\vec{a}$, which plays the role of an integer coordinate in the space of integrals; different sectors correspond to different patterns of positive and non-positive indices, and the reduction typically proceeds sector by sector. On this set of integrals one defines an ordering, typically based on some combination of the total degree $\sum_j a_j$, the number of positive indices, and additional tie-breaking rules (e.g. lexicographic order). This ranking is extended to equations by taking as the ``leading integral'' of an equation the highest-ranked integral appearing in it.

\item \emph{Generation and pruning of equations.} IBP identities are generated up to a given complexity, which typically produces a highly redundant set of equations. Many of them are linearly dependent.

    \item \emph{Removal of linearly dependent equations.} When an equation loses its leading integral due to previous substitutions, it may become identically zero or reduce to a relation among integrals that are already known in terms of lower-ranked ones. Such equations are recognized as linearly dependent and are dropped, which avoids unnecessary growth of intermediate expressions.

    \item \emph{Forward elimination.} The equations are processed in order of increasing complexity. For each equation, the highest-ranked integral on the left-hand side is chosen as a pivot and expressed in terms of lower-ranked integrals. This expression is then substituted into all other equations, gradually transforming the system into an (almost) triangular form with respect to the ranking.

    \item \emph{Back substitution.} Once forward elimination has produced relations for all non-master integrals in terms of lower-ranked ones, one performs a backward pass: starting from the lowest-ranked solved integrals, their expressions are recursively substituted into higher-ranked relations. This yields explicit reduction formulas in which every integral $I(\vec{a})$ is written as a linear combination of a finite set of master integrals with rational coefficients.
  \end{itemize}

Operationally, one starts from a finite set of seed integrals and generates IBP equations up to a given complexity, i.e. up to some maximal value of the ranking. The equations are then solved in such a way that higher-ranked integrals are expressed in terms of lower-ranked ones. When the system is sufficiently large, this elimination process stabilizes and one finds that all integrals in the family can be written as linear combinations
\begin{equation}
  I(\vec{a}) = \sum_{m} r_m(\vec{a}) \, I^{\text{master}}_m \,,
  \label{eq:IBP-reduction}
\end{equation}
where $I^{\text{master}}_m$ are the chosen master integrals and the coefficients $r_m(\vec{a})$ are rational functions in the kinematic invariants and $d$. The Laporta algorithm is purely algebraic: it does not rely on any particular representation of the integrals beyond the fact that they satisfy the IBP relations.

\subsection{Differential operators as ``integrals'' and Laporta-style reduction}
\label{subsec:derivative-laporta}

For a multivariate hypergeometric function $f(\vec{x})$ defined by a convergent hypergeometric series, we can derive a system of linear partial differential equations with rational coefficients,
\begin{equation}
  \mathcal{L}_\mu(\vec{x}, \partial_{\vec{x}}) \, f(\vec{x}) = 0 \,, 
  \qquad \mu = 1,\dots,M \,,
  \label{eq:hypergeom-PDE}
\end{equation}
where each $\mathcal{L}_\mu$ is a finite linear combination of monomials in the differential operators with coefficients rational in $\vec{x}$ and in the parameters. It is convenient to work with a fixed set of basic differential operators, for instance the Euler operators
\begin{equation}
  \theta_i = x_i \frac{\partial}{\partial x_i} \,, \qquad i = 1,\dots,n \,,
  \label{eq:Euler-ops}
\end{equation}
and to view all other differential operators as polynomials in the $\theta_i$ with coefficients rational in $\vec{x}$.

The analogy with the Laporta setup becomes apparent when we consider the infinite family of functions obtained by acting on $f$ with all monomials in the $\theta_i$:
\begin{equation}
  F_{\vec{\alpha}}(\vec{x}) \equiv \theta_1^{\alpha_1} \cdots \theta_n^{\alpha_n} f(\vec{x}) \,,
  \qquad \vec{\alpha} = (\alpha_1,\dots,\alpha_n) \in \mathbb{N}^n \,.
  \label{eq:derivative-family}
\end{equation}
Each multi-index $\vec{\alpha}$ now plays a role analogous to the set of propagator powers $\vec{a}$ in the Feynman-integral family \eqref{eq:Laporta-family}. Combinations $F_{\vec{\alpha}}$ act as ``integrals'' labelled by discrete data, and different choices of $\vec{\alpha}$ correspond to derivatives of different total order and distribution among the variables. In this language,

\begin{itemize}
  \item the \emph{combinations} of derivatives $F_{\vec{\alpha}}$ act as analogues of individual Feynman integrals, and
  \item the \emph{degrees} $\alpha_i$ act as analogues of the degrees (powers) of propagators.
\end{itemize}

The differential equations \eqref{eq:hypergeom-PDE} imply linear relations between the $F_{\vec{\alpha}}$. Writing
\begin{equation}
  \mathcal{L}_\mu(\vec{x}, \theta_1,\dots,\theta_n)
  = \sum_{\vec{\beta}} p_{\mu,\vec{\beta}}(\vec{x}) \,
    \theta_1^{\beta_1} \cdots \theta_n^{\beta_n} \,,
  \label{eq:Lmu-expansion}
\end{equation}
and applying $\mathcal{L}_\mu$ to $f$, we obtain
\begin{equation}
  0 = \mathcal{L}_\mu f 
    = \sum_{\vec{\beta}} p_{\mu,\vec{\beta}}(\vec{x}) \,
      F_{\vec{\beta}}(\vec{x}) \,.
  \label{eq:basic-relations}
\end{equation}
By further acting on these equations with additional differential operators, i.e. by considering
\begin{equation}
  \theta_1^{\gamma_1} \cdots \theta_n^{\gamma_n} \,
  \mathcal{L}_\mu(\vec{x},\theta) f(\vec{x}) = 0 \,,
  \label{eq:derived-relations}
\end{equation}
we generate an infinite collection of linear relations between the functions $F_{\vec{\alpha}}$. These relations are the direct analogue of the IBP relations \eqref{eq:IBP-relations} for Feynman integrals. An important difference is that there is no natural notion of \emph{sector} for hypergeometric functions: one does not partition the family into regions of the index space with different analytic properties. In this sense, the entire hypergeometric function (together with all its derivatives) behaves as a single sector, and the goal is to reduce the infinite family \eqref{eq:derivative-family} to a finite set of master combinations.

Once this analogy has been established, the Laporta idea can be applied literally in the space of derivatives. We proceed as follows:

\begin{enumerate}
  \item We introduce a ranking on the multi-indices $\vec{\alpha} \in \mathbb{N}^n$. In our implementation this ranking is slightly more structured than just using the total degree. To each multi-index
\[
  \vec{\alpha} = (\alpha_1,\dots,\alpha_n)
\]
we associate a weight vector
\[
  w(\vec{\alpha}) = \bigl( \max_i \alpha_i,\; \alpha_1 + \cdots + \alpha_n,\; \alpha_1,\dots,\alpha_n \bigr)\,,
\]
and we say that $\vec{\alpha} < \vec{\beta}$ if $w(\vec{\alpha})$ is lexicographically smaller than $w(\vec{\beta})$. Thus we first compare the largest component $\max_i \alpha_i$ (integrals/derivatives in which at least one direction has grown too much are considered ``more complicated''), then the total degree $\sum_i \alpha_i$, and only in case of a tie do we fall back to the usual lexicographic order on $(\alpha_1,\dots,\alpha_n)$. In this way we favor multi-indices where all components are relatively small and balanced, and we obtain a total order on the family $\{F_{\vec{\alpha}}\}$ that is well suited for elimination. 

This ordering was chosen heuristically and is by no means unique: in principle it can be replaced by any other reasonable total order on the multi-indices.  Empirically we find that in practice this ordering leads to somewhat more compact differential equations for the master functions.

  \item We choose a finite set of seed equations. Concretely, we take the original differential equations $\mathcal{L}_\mu f = 0$ and act on them with all monomials $\theta_1^{\gamma_1} \cdots \theta_n^{\gamma_n}$ up to some maximal total degree $|\vec{\gamma}| \leq D_{\text{max}}$. Each such operation produces an equation of the form
  \begin{equation}
    \sum_{\vec{\alpha}} q_{\vec{\alpha}}(\vec{x}) \, F_{\vec{\alpha}}(\vec{x}) = 0 \,,
  \end{equation}
  with coefficients $q_{\vec{\alpha}}(\vec{x})$ rational in $\vec{x}$ and in the parameters.

  \item We view all $F_{\vec{\alpha}}$ that appear in these equations as formal unknowns and rewrite the system as a linear algebra problem over the field of rational functions in $\vec{x}$ and in the parameters.

  \item We perform Gaussian elimination on this system, always solving for the highest-ranked $F_{\vec{\alpha}}$ in terms of lower-ranked ones. In contrast to typical IBP reductions for large Feynman-integral families, we do not implement a separate, dedicated step for identifying and removing linearly dependent equations. For the hypergeometric systems considered here, the number of genuinely redundant equations is usually small, and such equations are effectively eliminated during forward elimination: as pivots are chosen and substituted, linearly dependent relations either become identically zero or reduce to combinations of already-solved equations and can simply be discarded on the fly. This keeps the implementation conceptually simpler without noticeably affecting performance.
\end{enumerate}

If $D_{\text{max}}$ is chosen large enough, the elimination stabilizes and yields a finite set of master combinations $F_{\vec{\alpha}^{(1)}},\dots,F_{\vec{\alpha}^{(N)}}$ such that every other $F_{\vec{\alpha}}$ can be written as a linear combination
\begin{equation}
  F_{\vec{\alpha}}(\vec{x}) 
  = \sum_{m=1}^{N} R_{m,\vec{\alpha}}(\vec{x}) \, 
    F_{\vec{\alpha}^{(m)}}(\vec{x}) \,,
  \label{eq:derivative-reduction}
\end{equation}
with coefficients $R_{m,\vec{\alpha}}(\vec{x})$ rational in $\vec{x}$ and in the parameters. The functions
\begin{equation}
  \Phi_m(\vec{x}) \equiv F_{\vec{\alpha}^{(m)}}(\vec{x})
  = \theta_1^{\alpha^{(m)}_1} \cdots \theta_n^{\alpha^{(m)}_n} f(\vec{x}) \,,
  \qquad m=1,\dots,N \,,
\end{equation}
play the role of master integrals: they form a basis in terms of which all derivatives of $f$ can be expressed. In the next subsection we show how the reduction identities \eqref{eq:derivative-reduction} naturally give rise to a Pfaffian first-order system for the vector of master functions.

\subsection{Example function}
\label{subsec:H7ex}

As a concrete illustration of the reduction procedure described above, we now consider a specific bivariate hypergeometric function. This example is not intended to be special in any way; it simply provides a convenient test case that exhibits all essential features of our algorithm in a relatively compact setting.

As our example we choose the Horn function
\begin{equation}
H_7(\alpha ;\beta;\gamma;\delta;x,y) = \sum\limits_{m,n=0}^{\infty}\frac{(\alpha )_{2 m-n}(\beta )_n (\gamma )_n }{(\delta )_m}\frac{x^m y^n}{m!n!}
\label{eq:HornH7}
\end{equation}
defined by the double hypergeometric series in the variables $x$ and $y$. This function satisfies a holonomic system of partial differential equations, which we write in terms of the Euler operators $\theta_x = x \partial_x$ and $\theta_y = y \partial_y$.

The function $H_7$ in \eqref{eq:HornH7} obeys the following system of partial differential equations:
\begin{eqnarray}
\left[ (\delta -4 \alpha  x-2 x-1)\theta _x-\alpha  (\alpha +1) x+(1-4 x) \theta _x^2+(2 \alpha +1) x \theta _y
\right.
\nonumber \\
\left.+4 x \theta _x \theta
   _y-x \theta _y^2\right] H_7 & =& 0 ,
\nonumber \\
\left[2 \theta _x \theta _y+(\alpha -\beta  y-\gamma  y)\theta _y -\beta  \gamma  y-(y+1) \theta _y^2\right] H_7 & =& 0 .
\end{eqnarray}
Acting on these basic equations with additional powers of $\theta_x$ and $\theta_y$ we generate an infinite hierarchy of linear relations between higher derivatives of $H_7$. The first nontrivial members of this hierarchy read
\begin{eqnarray}
\left[\left(-\delta -\alpha ^2 x-\alpha  x+1\right)\theta _x +(\delta -4 \alpha  x-2 x-2)\theta _x^2 +(1-4 x) \theta _x^3
\right.
\nonumber \\
\left.
+(2
   \alpha +1) x \theta _x \theta _y+4 x \theta _x^2 \theta _y-x \theta _x \theta _y^2\right] H_7 & =& 0 ,
\nonumber \\
 \left[ (\alpha -\beta  y-\gamma  y)\theta _x \theta _y -\beta  \gamma  y \theta _x+2 \theta _x^2 \theta _y-(y+1) \theta _x \theta
   _y^2 \right] H_7 & =& 0 ,
   \nonumber \\
   \left[(\delta -4 \alpha  x-2 x-1)\theta _x \theta _y +(2 \alpha +1) x \theta _y^2-\alpha  (\alpha +1) x \theta _y
      \right.
\nonumber \\
\left.
   -x \theta _y^3
   +4 x
   \theta _x \theta _y^2+(1-4 x) \theta _x^2 \theta _y\right] H_7 & =& 0 ,
    \nonumber \\
    \dots
\end{eqnarray}
and so on. In principle this process can be continued indefinitely, producing a countable set of linear equations that relate all derivatives $\theta_x^{\alpha_1}\theta_y^{\alpha_2}H_7$.

To visualize the structure of this linear system it is convenient to represent it graphically. For this purpose we select a finite subset of equations (in this example we take 35 relations) and display them in matrix form, with the equations ordered along the vertical axis and the derivative monomials ordered along the horizontal axis. A red square at a given position indicates that the corresponding derivative enters the corresponding equation with a nonzero coefficient. The left panel of Fig.~\ref{fig:eqGraphical1} shows such an incidence diagram for the original system. In practice we use more equations than strictly necessary for the reduction; 35 equations are chosen here solely for visual clarity. For this particular function it is already sufficient to work with only 8 equations in order to derive a Pfaffian system.

Applying forward elimination in the sense of the Laporta algorithm transforms the system into an almost lower–triangular form, in which each equation is dominated by a single “pivot’’ derivative. Linear dependencies become manifest at this stage: equations that lose their leading derivative or become proportional to previous ones are discarded. For our illustrative choice of 35 equations, 5 such relations turn out to be linearly dependent and are dropped. The right panel of Fig.~\ref{fig:eqGraphical1} shows the resulting triangular structure after forward elimination, where only 30 independent equations remain.

\begin{figure}[h]
\centering
\begin{minipage}{0.45\textwidth}
\includegraphics[width=\textwidth]{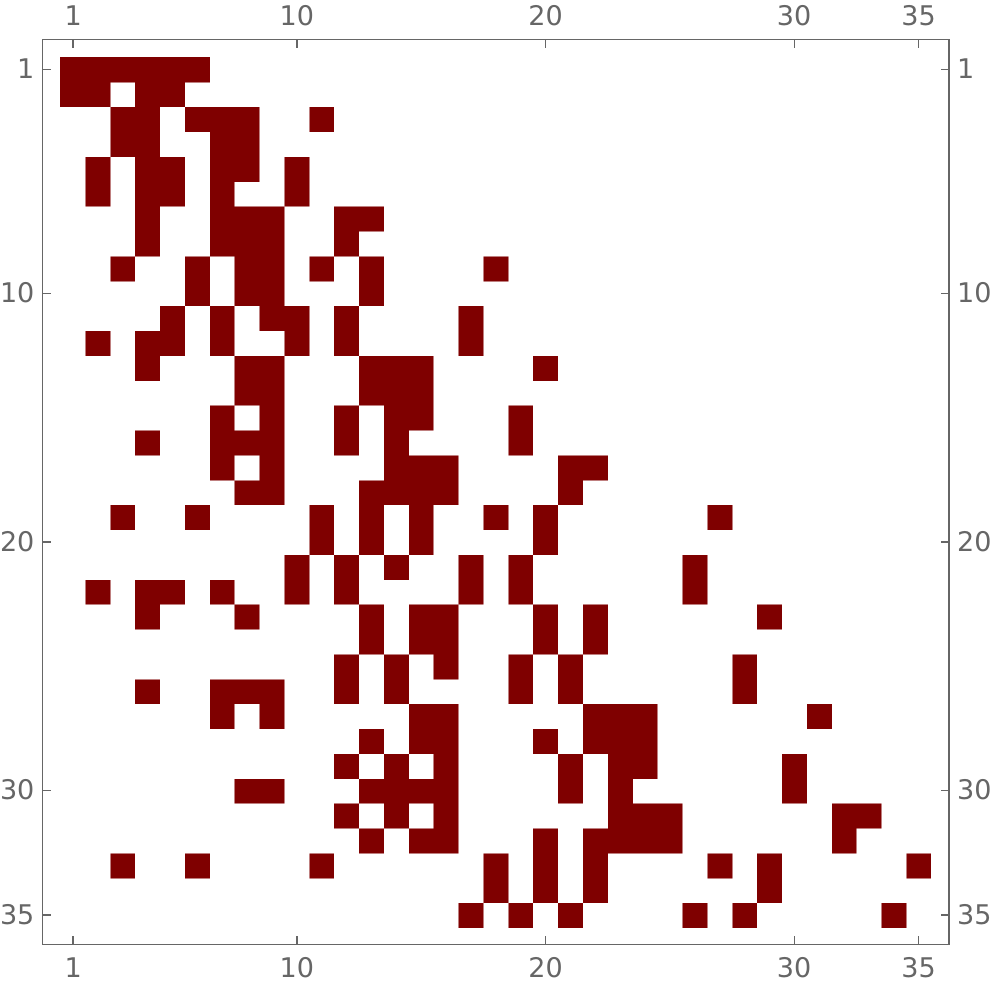}
\label{graph1}
\end{minipage}\hfill
\begin{minipage}{0.45\textwidth}
\includegraphics[width=\textwidth]{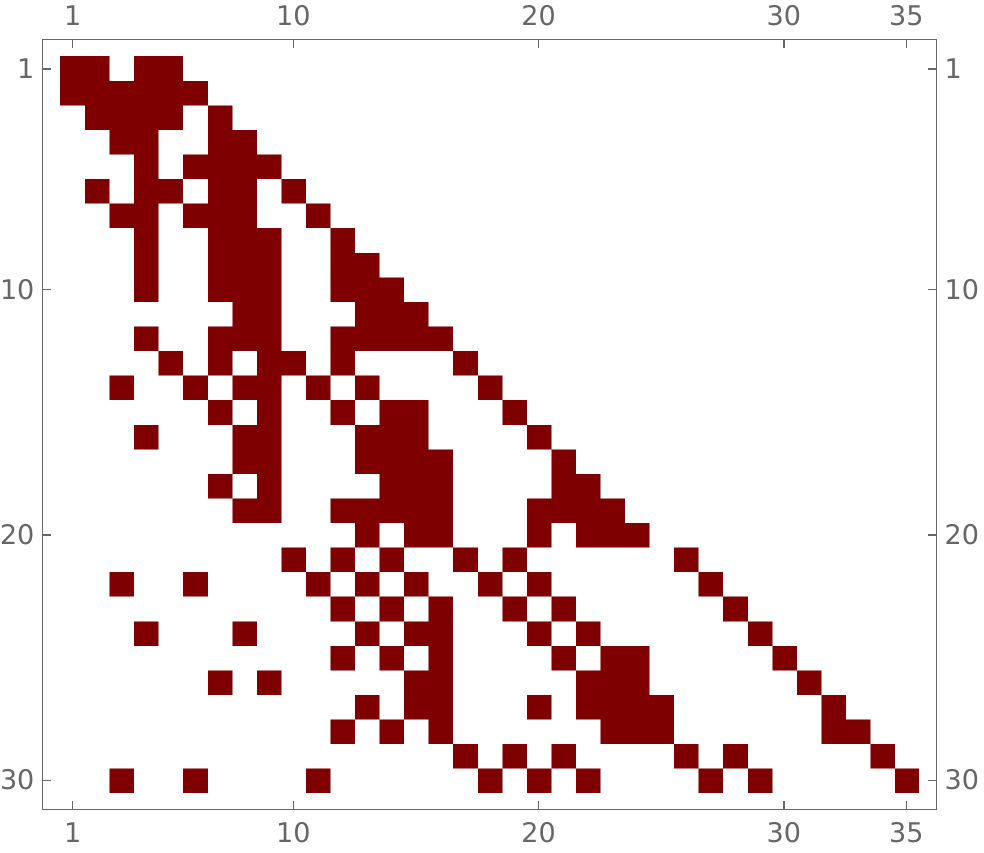}
\label{graph2}
\end{minipage}
\caption{Graphical representation of the linear system for the Horn function $H_7$. Left: incidence pattern of a set of 35 equations relating derivatives of $H_7$. Right: the same system after forward elimination, where the equations have been brought to (almost) lower–triangular form and 5 linearly dependent relations have been removed.}
\label{fig:eqGraphical1}
\end{figure}

\begin{figure}[h]
\centering
\begin{minipage}{0.45\textwidth}
\includegraphics[width=\textwidth]{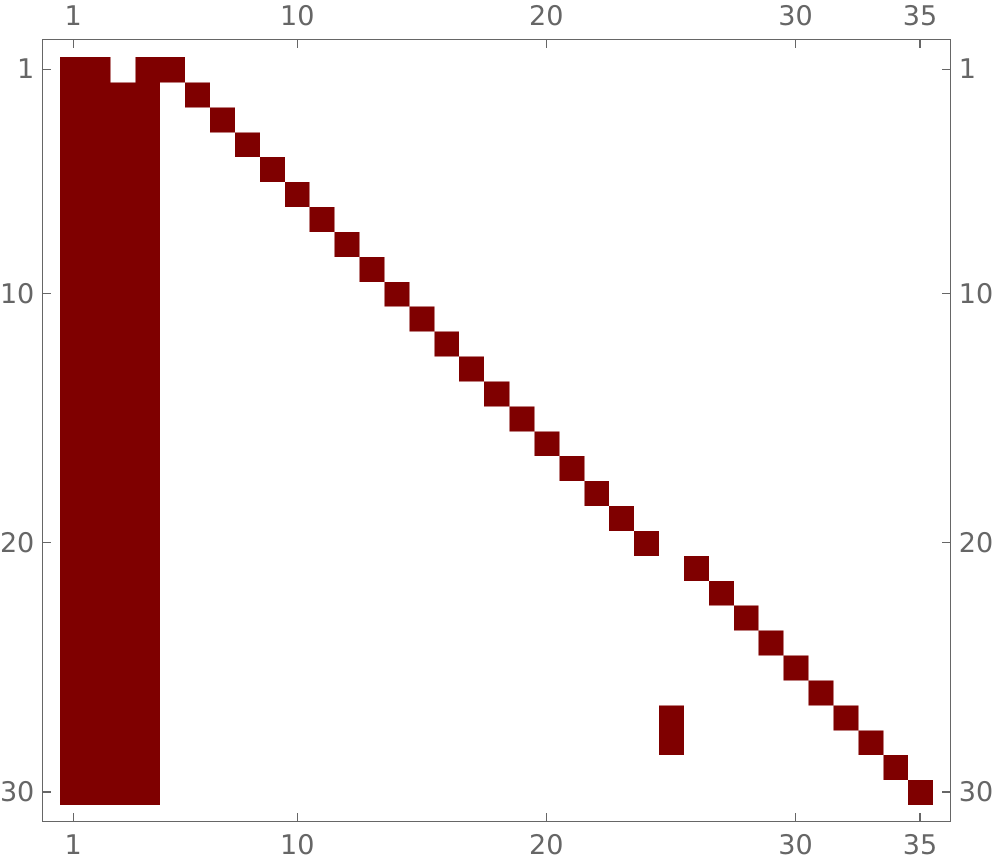}
\label{graph1}
\end{minipage}\hfill
\begin{minipage}{0.45\textwidth}
\includegraphics[width=\textwidth]{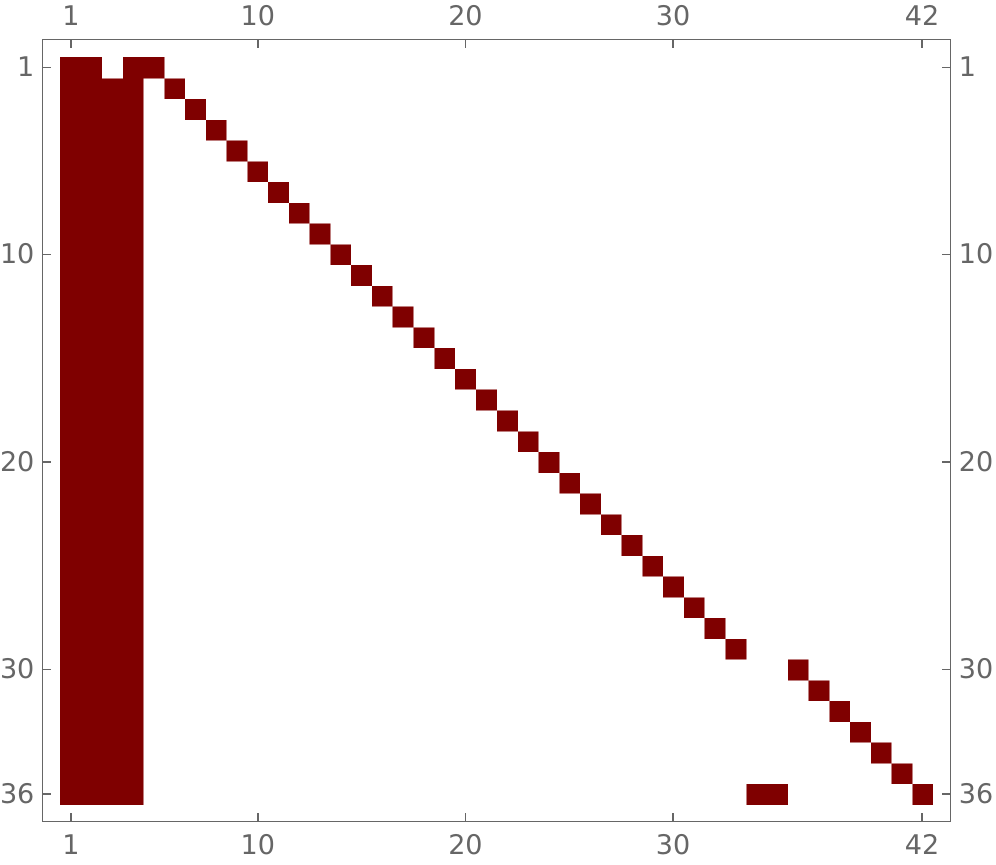}
\label{graph2}
\end{minipage}
\caption{Identification of master derivatives for the function $H_7$. Left: structure of the reduced system after back substitution when 35 equations are used; five master derivatives are visible. Right: analogous picture obtained from a system with 45 equations (9 are linearly dependent), illustrating how additional relations can eliminate spurious higher–order derivatives but may introduce other high–rank candidates. The true master set is identified by comparing several such reductions.}
\label{fig:eqGraphical2}
\end{figure}

The next step is back substitution. Starting from the highest–ranked derivatives we recursively express them in terms of lower–ranked ones, thereby identifying a finite set of master derivatives that cannot be eliminated. The result of this step is illustrated in Fig.~\ref{fig:eqGraphical2} (left panel), again using a graphical representation analogous to Fig.~\ref{fig:eqGraphical1}. From this picture it is clear that in our example we end up with five master derivatives, which we may choose as
\[
\{1,\theta_y,\theta_x,\theta_x\theta_y, \theta_x^4\theta_y^4 \}\,.
\]
The appearance of the derivative $\theta_x^4\theta_y^4$ in this set is somewhat artificial: it is an artefact of having included only a limited number of equations. If we generate more relations, this derivative can be expressed in terms of the lower–order ones.

This effect is illustrated in the right panel of Fig.~\ref{fig:eqGraphical2}, which shows the outcome of the same procedure when 45 equations are used from the outset. In that case the derivative $\theta_x^4\theta_y^4$ is no longer independent, but other, even higher derivatives appear as master candidates that cannot be reduced with the available system. In order to identify a stable set of true master derivatives, we therefore proceed as follows: we generate several systems of different (but sufficiently large) lengths, perform the full reduction for each of them, and then inspect which master derivatives remain unchanged across these systems. This stable core is taken as our final set of masters. All of these steps can be carried out with numerical values of the parameters $(x,y,\alpha,\beta,\gamma,\delta)$ substituted from the start, which greatly reduces the algebraic complexity and computation time at this exploratory stage.

In the case of $H_7$ this procedure leads us to the four–dimensional master basis
\begin{equation}
J = \{1,\theta_y,\theta_x,\theta_x\theta_y \}\,.
\end{equation}
Once this basis has been fixed, we no longer need to reduce all derivatives. Instead, in order to construct the Pfaffian system we only require expressions for the derivatives that appear on the right–hand side of the first–order equations for $J$, namely
\[
\left\{\theta _y^2,\theta _x^2,\theta _x \theta _y^2,\theta _x^2 \theta _y\right\}.
\]
For these derivatives we perform a second reduction step using a system of equations where the parameters $(x,y,\alpha,\beta,\gamma,\delta)$ are kept symbolic rather than numeric. Restricting the symbolic computation to this small subset of derivatives is sufficient to obtain the Pfaffian matrices and avoids the substantial overhead of a fully symbolic reduction of the entire family.

In this way we arrive at a Pfaffian system of the form
\begin{equation}
dJ = (M_x\, dx +M_y\, dy)J
\end{equation}
with
\begin{eqnarray}
M_x & =& \frac{A_0}{x}+\frac{A_{1/4}}{1-4x} +\frac{A_y}{4 x y^2-y^2-2 y-1}
\nonumber \\
M_y & =& \frac{B_0}{x}+\frac{B_{-1}}{1+y} +\frac{B_y}{4 x y^2-y^2-2 y-1}
\end{eqnarray}
where $4\times 4$ matrices $A_0,A_{1/4},A_y,B_0,B_{-1},B_y$ are rational functions of $(x,y)$ and the parameters. These matrices are too lengthy to be reproduced here, but they are straightforwardly generated by our reduction algorithm. One can verify that the integrability condition for a two–variable Pfaffian system,
\begin{equation}
\partial_y M_x -\partial_x M_y + M_xM_y - M_yM_x = 0\,,
\end{equation}
is satisfied identically. In addition, the correctness of the system can be checked by direct substitution of the defining series \eqref{eq:HornH7} into the differential equations, followed by expansion in $x$ and $y$ and comparison of coefficients. This confirms that the matrices $M_x$ and $M_y$ indeed provide a first–order representation of the holonomic system satisfied by the Horn function $H_7$.

\section{Generalized Frobenius method for Pfaffian systems}
\label{sec:frobenius}

In this section we briefly review the variant of the Frobenius method that we use for the numerical evaluation of multivariate hypergeometric functions on  their defining (any given) Riemann sheet\footnote{For the application of the Frobenius method in the context of Feynman diagrams, see for example Refs.~\cite{Frobenius1,Frobenius2,Frobenius3,Frobenius4,Frobenius5,KKOVelliptic2,Moriello:2019yhu,Bonciani:2019jyb,Frellesvig:2019byn,DiffExp,Bonisch:2021yfw, Bezuglov:2021tax, Blumlein:2021hbq,Bezuglov:2022npo,Armadillo:2022ugh,Liu:2022chg}.}. The general strategy follows our previous work on Lauricella functions \cite{Bezuglov:2025xol}.

We start from a first–order system in one complex variable $t$
\begin{equation}
  \frac{dJ}{dt} = M_t(t) \, J(t)\,,
  \label{eq:Pfaffian-ODE}
\end{equation}
obtained by restricting the full Pfaffian system
$dJ = \sum_i M_i(\vec{x})\,dx_i$ to a parametrized path in the space of variables. In practice we consider paths $t \mapsto \vec{x}(t)$ that connect a point where boundary data are known (typically close to the origin, where the defining hypergeometric series converges rapidly) to the target point of interest. The chosen path should not cross cuts and the latter are required to be fixed in $t$-plane, We choose them going parallel to real axis to $-\infty$.

Substituting $\vec{x}(t)$ into the matrices $M_i(\vec{x})$ yields a rational matrix function $M_t(t)$ with isolated singularities in the complex $t$–plane. We are thus reduced to solving \eqref{eq:Pfaffian-ODE} along a one–dimensional path.

The Frobenius method is applied in a neighbourhood of a regular singular point of \eqref{eq:Pfaffian-ODE}. After shifting the singularity to $t=0$ we can write
\begin{equation}
  M_t(t) = \frac{A_0}{t} + R(t)\,,
  \label{eq:M-expansion}
\end{equation}
where $A_0$ is the residue matrix at $t=0$ and $R(t)$ is a rational matrix function that is regular at the origin.\footnote{In the applications considered here the singularities encountered along the chosen paths are regular, or can be brought to regular form by a simple transformation; more singular behaviour is not required for the classes of hypergeometric functions we consider.} The object we construct is a matrix of fundamental solutions,
\begin{equation}
  \frac{dU}{dt} = M_t(t)\,U(t)\,, \qquad \det U(t) \neq 0\,,
  \label{eq:U-equation}
\end{equation}
whose columns form a basis of independent solutions. Once such a matrix is known at a given point $t$, the solution of \eqref{eq:Pfaffian-ODE} for any prescribed initial vector $J(t_0)$ is given by
\begin{equation}
  J(t) = U(t)\,U(t_0)^{-1} J(t_0)\,.
  \label{eq:J-from-U}
\end{equation}
In particular, if $t_0$ lies in a region where the original hypergeometric series converges and can be summed directly, \eqref{eq:J-from-U} provides a bridge between this region and the general point $t$.

The classical Frobenius method for a scalar equation seeks solutions of the form $t^\lambda$ times a power series, possibly multiplied by powers of $\ln t$ in resonant cases. For the matrix problem \eqref{eq:U-equation} we adopt an analogous ansatz, but we construct all independent solutions simultaneously in the form of a matrix series. Let $\lambda_1,\dots,\lambda_r$ be the eigenvalues of $A_0$ (possibly depending on external parameters such as an $\varepsilon$–expansion parameter). Eigenvalues whose differences are non–integers behave as in the non-resonant scalar case, whereas eigenvalues that differ by integers group into resonant blocks and lead to logarithmic terms in the Frobenius expansion. It is convenient to choose a representative exponent for each such block and to keep track of the possible powers of $\ln t$ separately.

We therefore consider an ansatz of the form
\begin{equation}
  U(t) = \sum_{\lambda \in S} t^{\lambda} \, U^{(\lambda)}(t)\,,
  \qquad
  U^{(\lambda)}(t) = \sum_{n=0}^{\infty} \sum_{k=0}^{m_\lambda}
    C^{(\lambda)}_{n,k}\, t^n (\ln t)^k\,,
  \label{eq:Frob-ansatz}
\end{equation}
where $S$ is a set of exponents obtained from the spectrum of $A_0$ by identifying and grouping resonant eigenvalues, and $m_\lambda$ is the maximal power of the logarithm associated with the block labelled by $\lambda$. The matrices $C^{(\lambda)}_{n,k}$ are constant (i.e.\ independent of $t$) and may depend analytically on parameters such as $\varepsilon$. Intuitively, each $\lambda \in S$ labels a “tower” of solutions with asymptotic behaviour $t^\lambda$ near the singular point, and the logarithmic factors encode the Jordan structure and resonances of $A_0$.
Substituting the ansatz \eqref{eq:Frob-ansatz} into the differential equation \eqref{eq:U-equation}, expanding $R(t)$ in a Taylor series around $t=0$, and matching powers of $t$ and $\ln t$, we obtain a nested system of linear recurrence relations for the coefficient matrices $C^{(\lambda)}_{n,k}$. 

In our implementation, the recurrence relations are not used merely to construct a local expansion around a single point. Instead, we embed the Frobenius series machinery into a global analytic-continuation procedure in the complex $t$–plane. Concretely, we cover the chosen continuation path by a chain of overlapping disks and, on each disk, represent the fundamental matrix $U(t)$ by a truncated Frobenius series. This turns the Frobenius method into both a provider of local solutions and a practical numerical “glue” that transports the solution from one disk to the next.

Let $\Sigma \subset \mathbb{C}$ denote the set of singular points of the one–dimensional system \eqref{eq:Pfaffian-ODE} (the images of the singular locus of the full Pfaffian system under the path parametrization $t \mapsto \vec{x}(t)$). We fix a path $P$ in the $t$–plane that connects an initial point $t_0 = 0$ inside the convergence domain of the defining hypergeometric series to the target point $t_\ast$. Around $t_0$ we construct a Frobenius expansion as in \eqref{eq:Frob-ansatz} and truncate it at some order $N$, obtaining a local representation $U^{[N]}_{(0)}(t)$ that is valid in a disk
\[
  D_0 = \{\, t \in \mathbb{C} \mid |t - t_0| < \rho_0 \,\},
\]
where the radius $\rho_0$ is chosen as the distance from $t_0$ to the nearest singularity,
\[
  \rho_0 = \min_{s \in \Sigma} |t_0 - s| \,,
\]
possibly multiplied by a safety factor $<1$ to account for numerical errors and to keep a margin away from the singular circle. Inside $D_0$ we can evaluate $U^{[N]}_{(0)}(t)$ and, via \eqref{eq:J-from-U}, obtain $J(t)$ for all $t$ on the portion of $\gamma$ contained in $D_0$.

\begin{figure}[ht]
	\centering
	\includegraphics[width=0.47\textwidth]{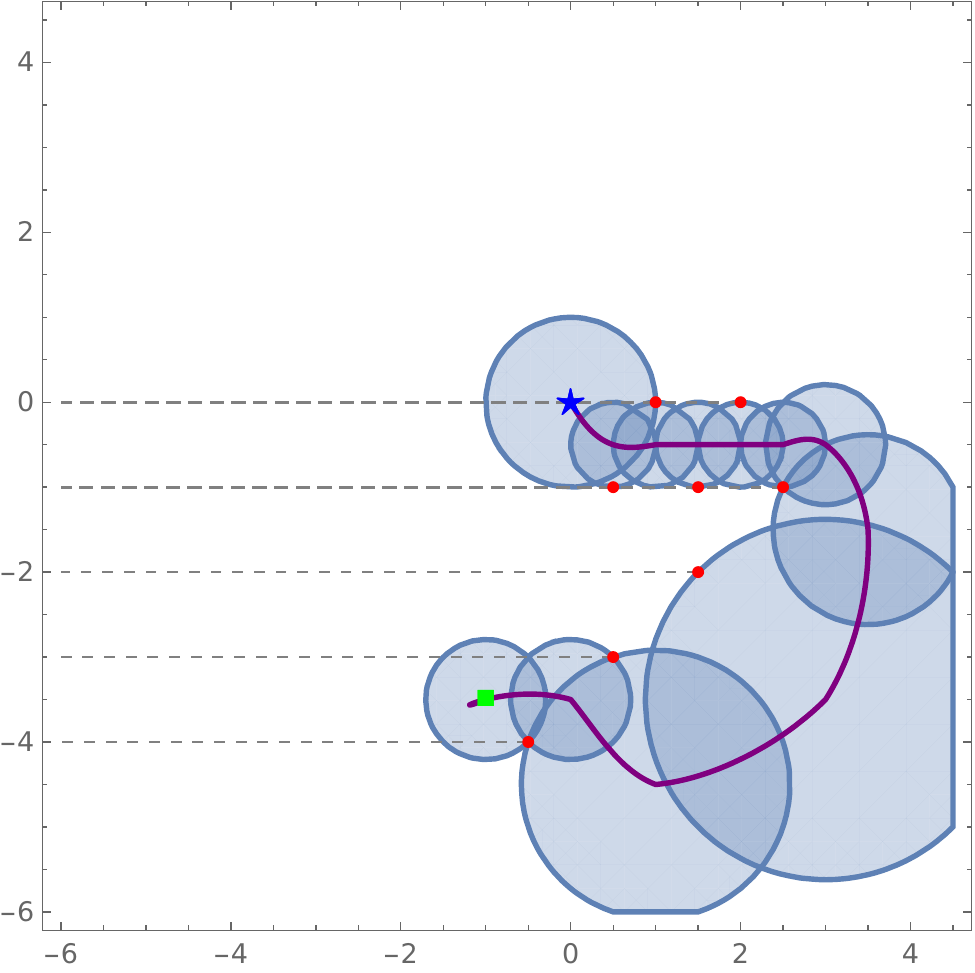}
	\caption{One of the possible paths with complex singular points is shown. The singular points are shown in red, the end point $t_\ast$ is marked by a green square, and the starting point $t_0$ by a blue star. The path of analytic continuation $P$ connecting them is indicated with a purple line. Shaded circles represent regions of convergence, and horizontal dotted lines indicate branch cuts of singular points.}
	\label{fig:path-disks}
\end{figure}

To proceed beyond $D_0$, we choose the next center $t_1$ further along the path $P$ so that the new disk around $t_1$ overlaps with the previous one, i.e. $D_0 \cap D_1 \neq \emptyset$. We then pick a matching point $t_{01}\in D_0\cap D_1$ and evaluate the truncated series $U^{[N]}_{(0)}(t_{01})$ to obtain numerical boundary data for a new Frobenius construction centered at $t_1$. Repeating the procedure, we determine the radius
\[
  \rho_1 = \min_{s \in \Sigma} |t_1 - s|
\]
and build a Frobenius series $U^{[N]}_{(1)}(t)$ valid in the disk
\[
  D_1=\{\,|t-t_1|<\rho_1\,\}.
\]

Iterating this step, we choose centers $t_j$ along $P$ such that consecutive disks overlap, $D_j\cap D_{j+1}\neq\emptyset$, yielding a chain of disks $D_0,D_1,\dots,D_M$ that covers the path from $t_0$ to $t_\ast$. The solution is propagated by analytic matching on each overlap: at stage $j$ we evaluate $U^{[N]}_{(j)}$ at a point $t_{j,j+1}\in D_j\cap D_{j+1}$ and fix the constant right factor relating the two local representations so that
\[
  U^{[N]}_{(j)}(t_{j,j+1}) = U^{[N]}_{(j+1)}(t_{j,j+1})\,C_{j+1}.
\]
This determines the continuation of the fundamental matrix to $D_{j+1}$, after which we advance to the next overlap. A schematic example of such an overlapping chain is shown in Fig.~\ref{fig:path-disks}, where the singularities, branch cuts, and the path $P$ are indicated explicitly.

Logarithmic terms in the Frobenius ansatz \eqref{eq:Frob-ansatz} are essential for capturing multivalued behavior. On each disk $D_j$, fixing a branch of $\ln(t-t_j)$ amounts to selecting a local Riemann sheet. Along the continuation path $P$ we therefore adopt a prescribed branch--cut pattern in the $t$--plane (e.g.\ radial cuts from the singular set $\Sigma$) and choose the disks and the path so as to avoid crossing these cuts. In \cite{Bezuglov:2025xol} we showed that one can always choose a continuation path that avoids crossing the prescribed branch cuts. In this case we remain on a fixed branch of the logarithms throughout the continuation.

From a numerical point of view, each disk $D_j$ comes with its own truncation order $N_j$, chosen so that the Frobenius series for $U^{[N_j]}_{(j)}(t)$ converges rapidly on the portion of $\gamma$ inside $D_j$. Since the radius $\rho_j$ is by construction limited by the nearest singularity, the ratio $|t - t_j|/\rho_j$ is uniformly bounded away from~1 along the path segment inside $D_j$, which ensures good convergence properties. The required orders $N_j$ can be estimated from simple bounds on the tail of the series, and we typically use a small safety margin to guarantee that the truncation error is well below the target precision. In this way, the generalized Frobenius method provides a systematic and controllable procedure to perform analytic continuation of the solution of the Pfaffian system along arbitrary paths in the complex plane, while keeping explicit track of singularities, branch cuts and Riemann sheets.

As an example of the described procedure, let us consider the evaluation of the following function:
\begin{equation}
H_7\left(\ep;\ep,\ep;\frac{1}{2}+\ep; 2, \frac{3}{2} \right)\,,
\end{equation}

Using the MultiHypExp \cite{Bera:2023pyz} software package, we can obtain the expansion of this function in terms of multiple polylogarithms,
\begin{eqnarray}
H_7\left(\ep;\ep,\ep;\frac{1}{2}+\ep; 2, \frac{3}{2} \right)&=&  1 -\left(G\left(0;2
   \sqrt{x}+1\right)+\frac{1}{2} G\left(1;\frac{4 \sqrt{x}}{2
   \sqrt{x}+1}\right)\right)\ep
   \nonumber\\
&&{} \left(-\frac{1}{2} G\left(1;\frac{4 \sqrt{x}}{2 \sqrt{x}+1}\right)
   G\left(-1;\left(2 \sqrt{x}+1\right) y\right)-G\left(0,-1;\left(2
   \sqrt{x}+1\right) y\right) \right.
      \nonumber\\&&{}
   -\frac{1}{2} G\left(1,\frac{2 \sqrt{x} y+y+1}{\left(2
   \sqrt{x}+1\right) y};\frac{4 \sqrt{x}}{2 \sqrt{x}+1}\right)+\frac{1}{2}
   G\left(0;2 \sqrt{x}+1\right)^2
\nonumber\\&&{}\left.
   +\frac{1}{2} G\left(1;\frac{4 \sqrt{x}}{2
   \sqrt{x}+1}\right) G\left(0;2 \sqrt{x}+1\right)\right)\ep^2 +
\mathcal{O}(\ep^3)\,,
\end{eqnarray}
and evaluate the latter numerically at the required point using tools like GiNaC \cite{Vollinga:2004sn} or HandyG \cite{Naterop:2019xaf}. This will provide us with an independent check of our calculation procedure.

The system of differential equations in Pfaffian form for the function $H_7$ was already found in the section~\ref{subsec:H7ex}.

Next, selecting the path $x \rightarrow 2 t$ and $y \rightarrow \frac{3t}{2}$, the $M_t$ differential system in the variable $t$ is defined by the matrix
\begin{equation}
M_t = \left(
\begin{array}{cccc}
 0 & \frac{1}{t} & \frac{1}{t} & 0 \\
 -\frac{3 \varepsilon ^2}{3 t+2} & -\frac{2 \varepsilon  (3 t-1)}{t (3 t+2)} & 0 & \frac{3
   (t+2)}{t (3 t+2)} \\
 -\frac{2 \varepsilon  (2 \varepsilon +3 t+2)}{(3 t+2) (8 t-1)} & \frac{2 (2 \varepsilon +12
   \varepsilon  t+3 t+2)}{(3 t+2) (8 t-1)} & -\frac{-2 \varepsilon +16 \varepsilon  t+8 t+1}{2 t
   (8 t-1)} & \frac{48 t^2+21 t-2}{t (3 t+2) (8 t-1)} \\
 -\frac{18 \varepsilon ^2 t (t+2) (2 \varepsilon +12 \varepsilon  t+3 t)}{(3 t+2) \left(72 t^3-9
   t^2-12 t-4\right)} & \frac{18 \varepsilon  t (t+2) (2 \varepsilon -24 \varepsilon  t-9
   t)}{(3 t+2) \left(72 t^3-9 t^2-12 t-4\right)} & -\frac{3 \varepsilon ^2 \left(48 t^2+21
   t-2\right)}{72 t^3-9 t^2-12 t-4} & \frac{S(t)}{2 t (3 t+2)
   \left(72 t^3-9 t^2-12 t-4\right)} \\
\end{array}
\right)\,,
\label{eq:MTEx}
\end{equation}
with 
\[
S(t) = 32 \varepsilon -2160 \varepsilon  t^4-216 t^4-2142
   \varepsilon  t^3-459 t^3+288 \varepsilon  t^2-90 t^2+168 \varepsilon  t-84 t-24
\]
and the boundary conditions vector by $b = \{1, 0, 0, 0\}^T$.
\begin{figure}[ht]
	\centering
	\includegraphics[width=0.5\textwidth]{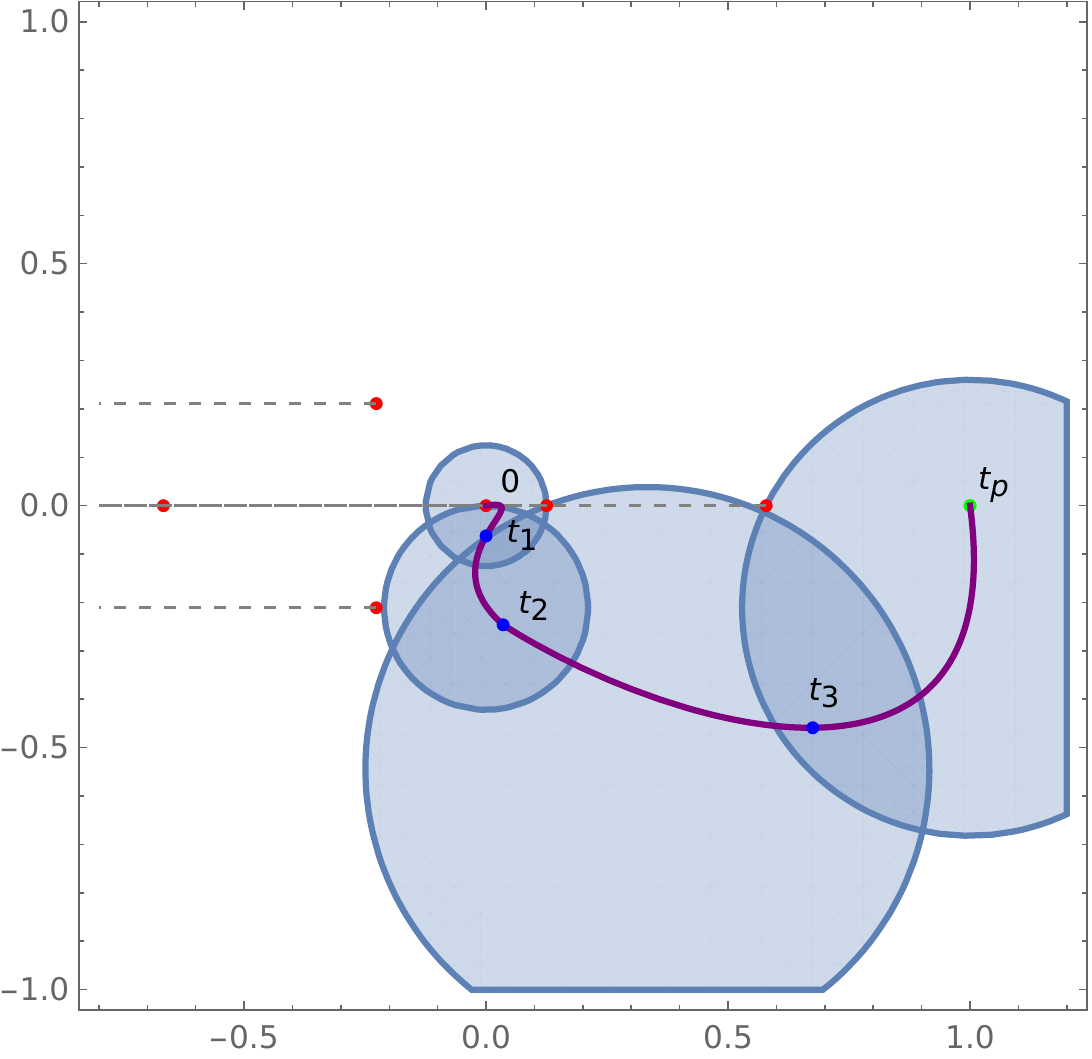}
	\caption{This figure illustrates the path of the analytic continuation of the system in Eq.~\eqref{eq:MTEx} from the origin $0$ to the evaluation point $t_p = 4$. Red points denote singularities, shaded circles represent convergence regions around the points $\left\{0,-\frac{27 i}{128},\frac{1}{3} - \frac{209 i}{384},1 - \frac{27 i}{128}\right\}$, blue points indicate the gluing points of these regions, horizontal dotted lines indicate branch cuts of singular points, the green point is our evaluation point and the path of analytic continuation $P$ connecting $0$ and $t_p$ is indicated by a purple line. }
	\label{fig:path}
\end{figure}
Thus, the problem reduces to solving this differential-equation system at the point $t_p = 1$. For that, we first need to construct an analytic-continuation path.  This path is not unique, our software package selects the path shown in Fig.~\ref{fig:path}. Consequently, the analytic continuation proceeds through the regions centered at the points $\left\{0,-\frac{27 i}{128},\frac{1}{3} - \frac{209 i}{384},1 - \frac{27 i}{128}\right\}$, and the value of the considered function at the evaluation point $t_p$ is assembled in the following way:
\begin{eqnarray}
H_7\left(\ep;\ep,\ep;\frac{1}{2}+\ep; 2, \frac{3}{2} \right)&=& P \left(U^{0}(0)\right)^{-1} b\, ,
\end{eqnarray}
where path $P$ is calculated as
\begin{eqnarray}
P &=&
U^{1 - \frac{27 i}{128}}(t_p) \left(U^{1 - \frac{27 i}{128}}(t_3)\right)^{-1}
U^{\frac{1}{3} - \frac{209 i}{384}}(t_3) \left(U^{\frac{1}{3} - \frac{209 i}{384}}(t_2)\right)^{-1}
\nonumber\\
&&{}\times
U^{-\frac{27 i}{128}}(t_2) \left(U^{-\frac{27 i}{128}}(t_1)\right)^{-1}
U^{0}(t_1)\, ,
\end{eqnarray}
 and $U^{S}(x)$ denotes the fundamental series solutions matrix obtained as a series expansion around the point $S$ and evaluated at the point $x$. The final step is the reconstruction of the $\ep$ expansion. For that, we use the Lagrange interpolation polynomials over small numerical values of $\ep$. For example, if we want the expansion up to $\ep^3$ with a precision of 30 decimal places, we can choose a set of four lattice points with a step size of $h = 10^{-18}$: $\ep \in \left\{-\frac{2}{10^{18}}, -\frac{1}{10^{18}}, \frac{1}{10^{18}}, \frac{2}{10^{18}}\right\}$. Notice that we specifically choose such small step size to achieve the required precision with some extra ``buffer" or ``stock," which generally allows for more reliable results. Combining everything together gives us the following result:
\begin{eqnarray}
  \lefteqn{H_7\left(\ep;\ep,\ep;\frac{1}{2}+\ep; 2, \frac{3}{2} \right)}
\nonumber\\  
&=&1
- (0.97295507452765665255267637172 + 1.57079632679489661923132169164 i) \varepsilon
\nonumber\\&&{}
+ (0.090439553877874924537407124413 + 0.523912153251492028016297851149 i) \varepsilon^2
\nonumber\\&&{}
- (7.46658998018013150687311766807 + 6.81276662646345510809702504262 i) \varepsilon^3
\nonumber\\&&{}
+\mathcal{O}(\varepsilon^4)\,.
\end{eqnarray}

\section{Practical applications}
\label{sec:aplications}

The algorithm described in this paper has many potential applications, most notably in the
numerical evaluation of multi-loop Feynman integrals. In particular, a wide class of integrals
can be represented in terms of multivariate hypergeometric functions of the type discussed
above, so that our Laporta–Frobenius framework can be used as a general-purpose numerical
engine for their evaluation and analytic continuation.

\begin{figure}[ht]
	\centering
	\includegraphics[width=0.83\textwidth]{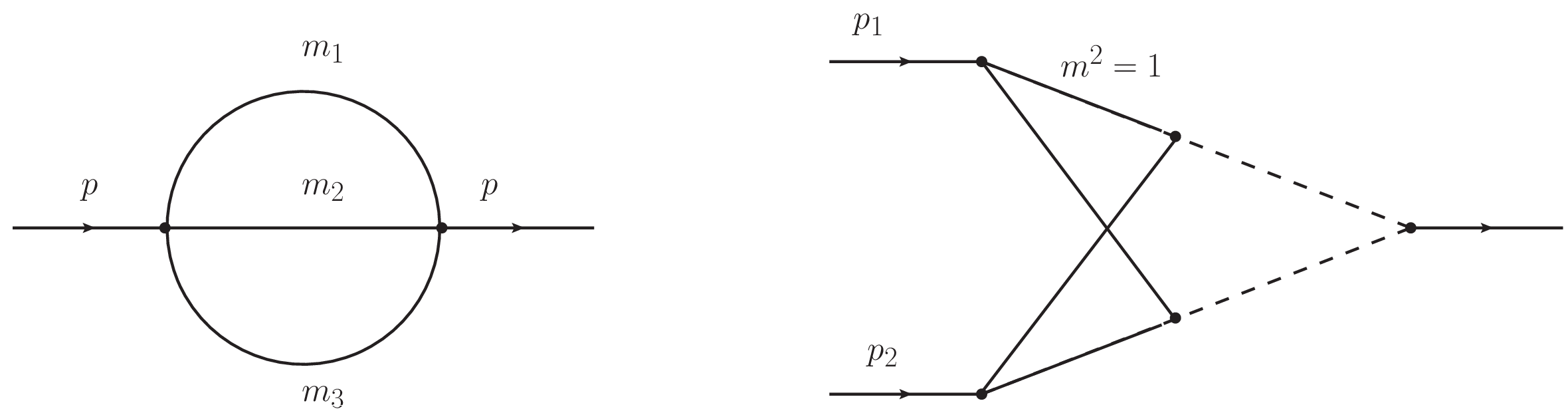}
	\caption{Examples of Feynman integrals expressible in terms of multivariate hypergeometric
	functions. Left: two-loop sunset diagram with three different internal masses. Right:
	non-planar two-loop vertex diagram with $p_1^2=p_2^2=0$ and $(p_1+p_2)^2 = 4s$; solid
	lines denote massive propagators and dashed lines massless ones.}
	\label{fig:examples}
\end{figure}

Some illustrative examples are shown in Fig.~\ref{fig:examples}. 

The simplest example in Fig.~\ref{fig:examples} is the massive sunset integral on the left. It can
be written in terms of Lauricella functions $F_C^{(3)}$ of three variables as
\cite{Berends:1993ee, Ananthanarayan:2019icl}
\begin{align*}
S_1 = &-m_3^2\left( \frac{m_3^2}{4\pi}\right) ^{-2\varepsilon}
\\
&\times \Big\{
z_1^{1-\varepsilon} z_2^{1-\varepsilon}\,\Gamma ^2(\varepsilon-1)\,
F_C^{(3)}\!\left(1,\, 2-\varepsilon;\, 2-\varepsilon,\, 2-\varepsilon,\, 2-\varepsilon;\, z_1, z_2, z_3\right)
\\
&\qquad - z_1^{1-\varepsilon}\,\Gamma ^2(\varepsilon-1)\,
F_C^{(3)}\!\left(1,\, \varepsilon;\, 2-\varepsilon,\, \varepsilon,\, 2-\varepsilon;\, z_1, z_2, z_3\right)
\\
&\qquad - z_2^{1-\varepsilon}\,\Gamma ^2(\varepsilon-1)\,
F_C^{(3)}\!\left(1,\, \varepsilon;\, \varepsilon,\, 2-\varepsilon,\, 2-\varepsilon;\, z_1, z_2, z_3\right)
\\
&\qquad - \Gamma(1-\varepsilon)\,\Gamma(\varepsilon-1)\,\Gamma(2\varepsilon-1)\,
F_C^{(3)}\!\left(2\varepsilon-1,\, \varepsilon;\, \varepsilon,\, \varepsilon,\, 2-\varepsilon;\, z_1, z_2, z_3\right)
\Big\},
\end{align*}
where $z_1 = \frac{m_1^2}{m_3^2}$, $z_2 = \frac{m_2^2}{m_3^2}$, $z_3 = \frac{p^2}{m_3^2}$ and
\[
F_C^{(3)}(\alpha;\beta;\gamma_1,\gamma_2,\gamma_3;x,y,z) =
\sum\limits_{m,n,p=0}^{\infty}\frac{(\alpha)_{m+n+p} (\beta)_{m+n+p}}{(\gamma_1
   )_{m}(\gamma_2
   )_{n}(\gamma_3
   )_{p}}\frac{x^m y^n z^p}{m!n!p!}\,.
\]
Our algorithm can be used to perform the analytic continuation of these Lauricella functions
to arbitrary kinematic configurations. For example, at
$m_1^2 = 3$, $m_2^2=2$, $m_3^2 = 1$ and $p^2 = 4$ we obtain the following
$\ep$–expansion:
\begin{align}
S_1 &= \frac{3}{\ep^2}
      - \frac{0.1454252165334168566592724942}{\ep}
      + 11.6783303708533400509414315111
\notag\\
&\quad
      + 4.086774082457494130339365088\,\ep
      + 37.967852023599798748163023286\,\ep^2\,.
\end{align}

A more involved example is the non-planar two-loop elliptic vertex shown on the right-hand
side of Fig.~\ref{fig:examples}. As demonstrated in Ref.~\cite{Bezuglov:2021tax}, one of its
master integrals $N_1$ can be expressed in terms of generalized Kampé de Fériet functions
$\Fkdf{p:q:k}{l:m:n}{\dots}{\dots}{\dots}{\dots}{\dots}{\dots}{x ; y}$ and generalized hypergeometric
functions:
\begin{align}
N_1=  &
-\frac{\pi ^{3/2}  (\varepsilon+6) \text{csc} (\pi  \varepsilon ) \Gamma (2 \varepsilon+2)}{2^{2 \varepsilon +3}(\varepsilon +2) \Gamma (3-\varepsilon )
\Gamma \left(\varepsilon +\frac{5}{2}\right)}\times \nonumber\\
&\times\Bigg\{ \Fkdf{2:1:5}{2:0:4}{2, 2(1+\ep)}{\frac{5}{2},\frac{5}{2}+\ep}{1}{-}{1,2,\frac{5}{2},2+\ep,3+\frac{\ep}{3}}{3,3-\ep,3+\ep,2-\frac{\ep}{3}}{-\frac{s}{4} ; -s}\nonumber\\
&-\frac{\varepsilon  (\varepsilon +1) (3 \varepsilon+10)}{2 (\varepsilon +6)} \Fkdf{2:1:6}{2:0:5}{2, 2(1+\ep)}{\frac{5}{2},\frac{5}{2}+\ep}{1}{-}{1,2,\frac{5}{2},2+\ep,2+\ep,3+\frac{3\ep}{5}}{3,3-\ep,3+\ep,3,2+\frac{3\ep}{5}}{-\frac{s}{4} ; s}\Bigg\}\nonumber\\
&+\frac{\pi \cot (\pi  \varepsilon ) \Gamma (2 \varepsilon+3) }{3  (-s)^{\varepsilon } 4^{\varepsilon +1} (1-4 \varepsilon ^2)}
\Fkdf{2:1:5}{2:0:4}{2+\ep, 2-\ep}{\frac{5}{2},\frac{5}{2}-\ep}{1}{-}{1,1,1-\ep,\frac{3}{2}-\ep,2-\frac{2\ep}{3}}{2,2-2\ep,2-\ep,1-\frac{2\ep}{3}}{-\frac{s}{4} ; -s}\nonumber\\
&+\frac{\pi ^{3/2} 2^{-2 \varepsilon -3} (\varepsilon+6) \text{csc} (\pi  \varepsilon ) \Gamma (2 \varepsilon+2)}{(\varepsilon +2) \Gamma (3-\varepsilon )\Gamma \left(\varepsilon +\frac{5}{2}\right)}\pFq{6}{5}{1,2,2,2(1+\ep),2+\ep,3+\frac{\ep}{3}}{3,3-\ep,3+\ep,\frac{5}{2}+\ep,2+\frac{\ep}{3}}{-s}
\nonumber\\
&-\frac{\pi  \varepsilon  (3 \varepsilon +10) \text{csc} (\pi\varepsilon ) \Gamma (\varepsilon +2)}{4 \left(2\varepsilon ^2+7 \varepsilon +6\right) \Gamma(3-\varepsilon )}\pFq{7}{6}{1,2,2,2(1+\ep),2+\ep,2+\ep,3+\frac{3\ep}{5}}{3,3,3-\ep,3+\ep,\frac{5}{2}+\ep,2+\frac{3\ep}{5}}{s}
\nonumber\\
&-\frac{\pi ^{3/2} 2^{-2 (\varepsilon +2)} \text{csc} (\pi \varepsilon ) \Gamma (2 \varepsilon +3)}{\Gamma(2-\varepsilon ) \Gamma \left(\varepsilon+\frac{5}{2}\right)}\pFq{3}{2}{1,2,2(1+\ep)}{\frac{5}{2},\frac{5}{2}+\ep}{-\frac{s}{4}}\,,
\end{align}
where the generalized Kampé de Fériet functions are defined by \cite{MR834385}
\begin{equation}
\Fkdf{p:q:k}{l:m:n}{(a_p)}{(\alpha_l)}{(b_q)}{(\beta_m)}{(c_k)}{(\gamma_n)}{x ; y} =
\sum\limits_{r,s = 0}^{\infty}\frac{\prod\limits_{j=1}^p(a_j)_{r+s}\prod\limits_{j=1}^q(b_j)_{r}\prod\limits_{j=1}^k(c_j)_{s}}{\prod\limits_{j=1}^l(\alpha_j)_{r+s}\prod\limits_{j=1}^m(\beta_j)_{r}\prod\limits_{j=1}^n(\gamma_j)_{s}}\frac{x^r}{r!}\frac{y^s}{s!}.
\end{equation}
Note that all hypergeometric functions in this expression satisfy condition \eqref{eq:condition}.

Our method can again be used to perform the analytic continuation of all hypergeometric
functions appearing in $N_1$. For example, at $s=3$ we find
\begin{align}
N_1 &= -0.1552055907861132835488885437
      + 0.1294388500389800147114802077\, i
\notag\\
&\quad
      -\left(0.235240853378471909605144342
             + 0.4220368751015038627780374179\, i\right)\ep
\notag\\
&\quad
      +\left(0.371797477245885038452548728
             - 0.03921305677983090697281778\, i\right)\ep^2\,.
\end{align}
When obtaining such expressions it is crucial to keep track of the branches of all multivalued
building blocks, in particular $(-s)^{\ep}$ and the generalized hypergeometric function
$\pFq{7}{6}{1,2,\dots}{3,\dots}{s}$. Different choices of branches lead to different numerical values
for $N_1$; in Sec.~\ref{sec:monodromies} we described how these values are related by monodromy
transformations and how our framework makes this structure explicit.

Of course, the scope of our method is not limited to these two examples. It can equally well be
applied to series obtained from GKZ-type integral representations \cite{GKZ1,GKZ2,GKZ3,GKZ4,GKZ5,beukers2013monodromy,Kalmykov:2012rr,delaCruz:2019skx,Klausen:2019hrg,Ananthanarayan:2022ntm}, as well as to other families
of multi-loop integrals. For instance, it is known that “watermelon’’ (or “banana’’) integrals with
an arbitrary number of loops can always be expressed in terms of Lauricella functions
$F_C^{(n)}$ \cite{Lee:2019lsr}; these fall directly within the range of our algorithm. We therefore
expect the techniques developed here to become a useful tool for the high-precision evaluation
and analytic study of a wide variety of Feynman integrals and related special functions.

Multivariate hypergeometric functions also appear in many other areas of physics and
mathematics. A systematic review of this literature would take us far beyond the scope of the
present work, but it is useful to mention a few representative examples. In the context of
conformal field theory, Ref.~\cite{Li:2019dix} derives closed-form expressions for crossed-channel
(t-channel) conformal blocks in the lightcone expansion, using the spin-pole structure implied by
the Lorentzian inversion formula. The final result is written in terms of Kampé de Fériet
two-variable hypergeometric functions, in particular functions of the form
\[
\Fkdf{2:0:2}{1:0:2}{A_3, A_4}{B_3}{-}{-}{A_1, A_2}{B_1, B_2}{x ; y},
\]
which fit naturally into the general hypergeometric framework treated in this paper.

Hypergeometric functions of three and more variables also appear explicitly in several other
settings. For instance, Lauricella functions of type $F_D^{(4)}$ arise in problems in general
relativity~\cite{cartin2006wave}, in models of fluctuating radio channels and related
radiophysics applications~\cite{al2023fluctuating,sanchez2023multi}, and in the study of
Seiberg–Witten theory~\cite{Akerblom:2004cg}. In all these cases the relevant special functions already appear in the form of multivariate hypergeometric series of the types discussed above, and are therefore directly accessible to the numerical methods developed in this work.

Taken together, these examples indicate that our approach and the accompanying \texttt{HAPC}
package may find practical use in a wide range of problems beyond multi-loop Feynman
integrals. They may also motivate further work in expressing complicated analytic results
directly in terms of multivariate hypergeometric functions, rather than ad hoc integral
representations. To the best of our knowledge, no general-purpose algorithm for the numerical
evaluation and analytic continuation of such functions was previously available; typically each
individual function had to be treated separately. The present work is a step towards filling this
gap by providing a unified and systematically extensible framework.

\section{Monodromies}
\label{sec:monodromies}

Multivariate hypergeometric functions are, in general, multivalued analytic functions. Their values at a given point depend not only on the coordinates of that point but also on the homotopy class of the path along which the function is analytically continued from some fixed base point. Different choices of paths, winding in different ways around the singular locus, lead to different points on the Riemann surface of the function. As a consequence, different implementations of “the same’’ hypergeometric function in computer algebra systems may return different complex numbers at the same argument, depending on how the principal branch is defined internally.

As a simple motivation, consider the Appell function
\begin{equation}
    F_3\left(1,\frac{1}{2};\frac{1}{3},\frac{1}{4};\frac{1}{5};2-i,3+i\right).
    \label{eq:F3functionExample}
\end{equation}
If this value is computed in \textit{Wolfram Mathematica v.14.3.0}, one obtains
\[
F_3\left(1,\tfrac{1}{2};\tfrac{1}{3},\tfrac{1}{4};\tfrac{1}{5};2-i,3+i\right)
\approx -3.77866 + 3.94349\, i,
\]
whereas \textit{Maple 2024} returns
\[
F_3\left(1,\tfrac{1}{2};\tfrac{1}{3},\tfrac{1}{4};\tfrac{1}{5};2-i,3+i\right)
\approx -1.81221 + 0.3426\, i.
\]
This discrepancy does not indicate a bug in either program: both values are correct, but they lie on different Riemann sheets and correspond to different conventions for choosing the principal value and for placing branch cuts in the complex domain.

It is important to note that the Riemann surface itself obtained after gluing all Riemann sheets along the cut lines is the same for any of the approaches and does not depend on how the cuts are defined.

The aim of this section is to show that the Frobenius-based approach developed in this paper can be used to construct analytical continuations of hypergeometric functions with the account of multiple Riemann sheets. In subsections 
\ref{subsec:monodromy-1d} and \ref{subsec:monodromy-2d} we first show monodromy calculations for simple hypergeometric functions in one and two variables without paying attention to different Riemann sheets and cuts involved. Working with the Pfaffian system satisfied by $F_3$, we construct monodromy matrices that describe the effect of encircling components of the singular locus on a basis of local solutions. In this framework, the different values returned by \textit{Mathematica} and \textit{Maple} can be connected by explicit monodromy transformations, and, more generally, we can systematically enumerate all possible values of a given multivariate hypergeometric function at a specified point by following paths in different homotopy classes. Then in subsection \ref{subsec:monodromy-computation} we give additional details on monodromy calculation within Frobenius method and finally in \ref{subsec:Brunch} we present a systematic procedure for analytic continuation of hypergeometric functions across multiple Riemann sheets.

\subsection{Elementary example}\label{subsec:monodromy-1d}

To illustrate these ideas in the simplest possible setting, it is useful to start with a hypergeometric function that can be written explicitly in terms of elementary functions. Consider
\begin{equation}
\, _2F_1\left(\frac{1}{3},1;2;x\right) = \frac{3 \left(1-(1-x)^{2/3}\right)}{2 x} \,.
\end{equation}
The right–hand side involves a fractional power of $1-x$, so the function is manifestly multivalued: the cubic root has three different branches. As a consequence, evaluating $_2F_1\big(\tfrac{1}{3},1;2;x\big)$ at a point such as $x = 2 + i$ does not give a single value, but three distinct values corresponding to the three branches of $(1-x)^{2/3}$. Numerically one obtains
\[
\, _2F_1\left(\frac{1}{3},1;2;2+i\right) \in 
\left\{
\begin{aligned}
  &0.977976 + 0.455953\, i\,,\\
  &1.06569 - 1.00531\, i\,,\\
  &-0.243662 - 0.350639\, i
\end{aligned}
\right\}.
\]
Our goal in this subsection is to show how these three values can be obtained in a systematic way using the Frobenius-based approach to analytic continuation, rather than by explicitly manipulating branches of the elementary representation.

We introduce a two–dimensional basis of solutions built from the function and its Euler derivative,
\[
J=\{1,\theta_x\}\,{}_2F_1\left(\frac{1}{3},1;2;x\right),
\]
and write the corresponding first–order system in the form
\begin{equation}
\partial_x J = \left(
\begin{array}{cc}
 0 & \frac{1}{x} \\
 -\frac{1}{3 (x-1)} & -\frac{4 x-3}{3 (x-1) x} \\
\end{array}
\right) J \,.
\end{equation}
This is a scalar second–order hypergeometric equation rewritten as a $2\times 2$ Pfaffian system with regular singular points at $x=0$, $x=1$ and $x=\infty$. Once the system is in this form, all information about the different branches of the solution is encoded in the monodromy of the fundamental matrix along paths in the complex $x$–plane.

\begin{figure}[ht]
	\centering
	\includegraphics[width=0.5\textwidth]{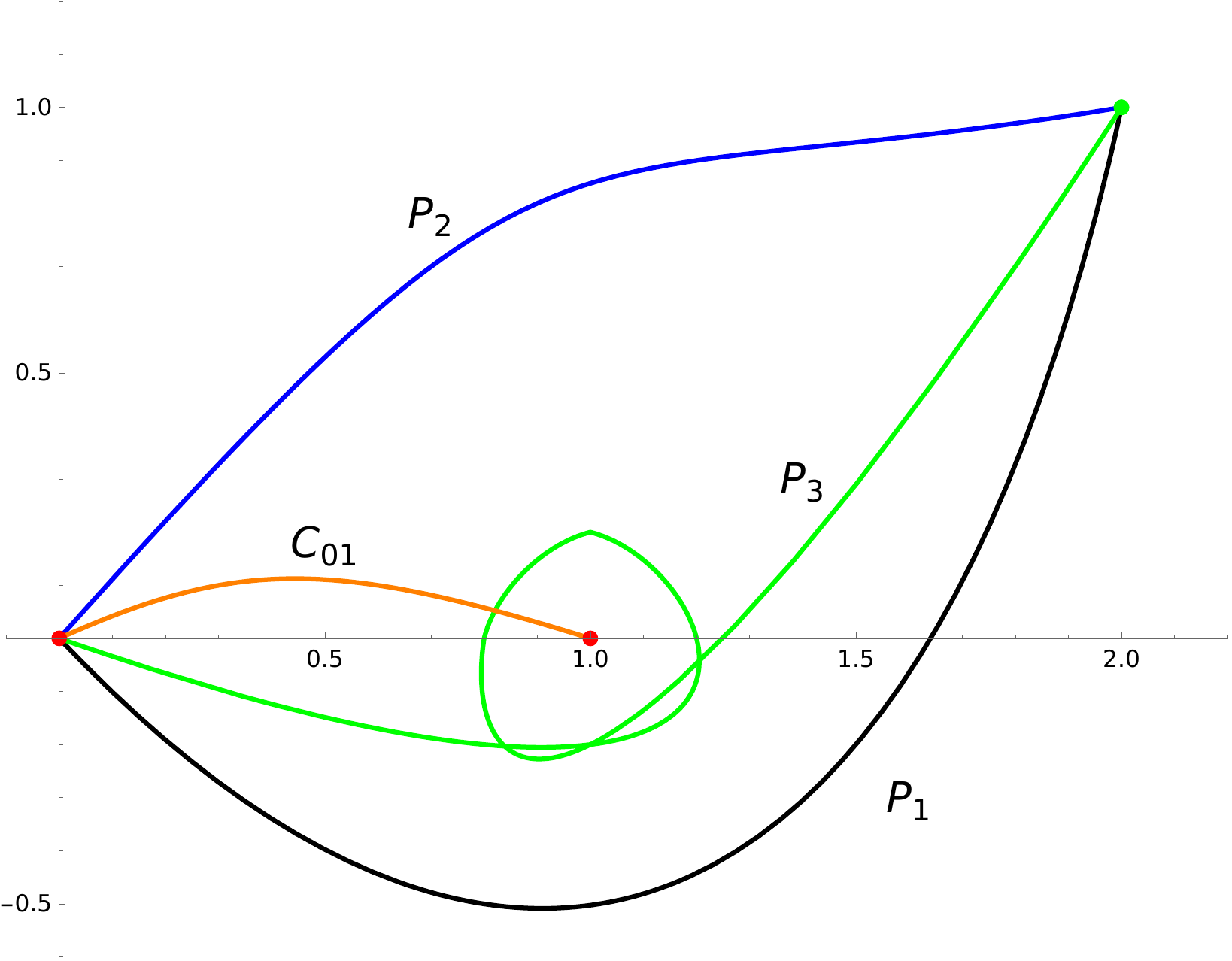}
	\caption{Schematic picture of analytic continuation in the $x$–plane for the elementary example. The red dots indicate the singular points of the hypergeometric equation at $x=0$ and $x=1$, while the green dot marks the target point $x = 2+i$ where the function is evaluated. The coloured curves show three representative paths $P_1$, $P_2$ and $P_3$ from the common base point at $x=0$ to $x=2+i$, which differ only by how they wind around the singularity at $x=1$. Each homotopy class of such paths corresponds to a different branch of the multivalued function $_2F_1\!\left(\tfrac{1}{3},1;2;x\right)$ at the target point.}
	\label{fig:paths-elem}
\end{figure}

We can obtain all three branches of $_2F_1\big(\tfrac{1}{3},1;2;x\big)$ at $x=2+i$ by solving this differential equation along different paths from a fixed base point $x_0$ (for instance, $x_0 = 0$) to the target point $x=2+i$. The particular value of the function corresponding to a given solution is determined entirely by the homotopy class of the path in the complex plane with the singular points removed. Three such paths, denoted $P_1$, $P_2$ and $P_3$, are shown schematically in Fig.~\ref{fig:paths-elem}. They all start at the same base point and end at $x=2+i$, but differ in how they wind around the singular point $x=1$. In this sense they belong to different homotopy classes in the punctured plane $\mathbb{C} \setminus \{0,1\}$.

Given a fundamental matrix $U(x)$ normalized at $x_0$, the parallel transport along a path $P$ from $x_0$ to $x$ can be written as a path–ordered product of local evolution operators. Acting on a fixed boundary vector $J(x_0)$ (determined, for instance, by the Taylor series of the hypergeometric function at $x=0$), this produces the value of $J(x)$ at the endpoint of the path. Different paths with the same endpoints but different homotopy classes yield different values of $J(x)$ and hence different branches of $_2F_1$.

The effect of changing the path is conveniently described in terms of monodromy matrices. The monodromy around a singular point $x_s$ is the linear transformation obtained by analytically continuing the fundamental matrix $U(x)$ once around a small loop encircling $x_s$ in the positive (counterclockwise) direction. This transformation is represented by a matrix $M_s$ acting on $U(x)$ from the right, and its inverse $M_s^{-1}$ corresponds to a loop in the opposite (clockwise) direction. In our example we are particularly interested in the monodromy around $x=1$, which we denote by $M_1$. In a convenient basis this matrix takes the form
\begin{equation}
M_1 = \left(
\begin{array}{cc}
 1 & 0 \\
 -\frac{3}{2}-\frac{i \sqrt{3}}{2} & -\sqrt[3]{-1} \\
\end{array}
\right).
\end{equation}
One can verify that
\[
M_1^3 = 1,
\]
which reflects the fact that a triple loop around $x=1$ brings the function back to the original branch. (Strictly speaking, the full monodromy group is generated by loops around $x=0$ and $x=1$, but in the present example the monodromy around $x=0$ is the identity, so only $M_1$ is nontrivial.) The subgroup generated by $M_1$ is thus cyclic of order three and is naturally isomorphic to $\mathbb{Z}/3\mathbb{Z}$, in agreement with the presence of three distinct branches of the function.

Monodromy matrices allow us to relate the solutions obtained along different paths. However, we cannot apply $M_1$ directly in the basis normalized at $x_0=0$, because none of the paths $P_1$, $P_2$ or $P_3$ passes exactly through $x=1$. Instead, we introduce an auxiliary path $C_{01}$ from $x_0=0$ to a point near $x=1$, see Fig.~\ref{fig:paths-elem}. This path is computed in exactly the same way as any other path in our Frobenius framework: we cover it by overlapping disks, construct local Frobenius expansions in each disk, and multiply the corresponding local evolution matrices along the path. Schematically we write
\[
C_{01} = U^{1}(1/2)\,U^{0}(1/2),
\]
where $U^{0}$ is the fundamental matrix normalized at $x=0$ and $U^{1}$ is the one normalized at $x=1$; the point $x=1/2$ serves as an overlap point where the two local descriptions can be matched. Simply put, the matrix $C_{01}$ connects the neighborhood of point 0 with the neighborhood of point 1. The neighborhood is considered where the Frobenius series converges, that is, to the nearest singularity.

Using $C_{01}$, we define a conjugated monodromy matrix
\begin{equation}
\bar{M}_1 = C_{01}^{-1} M_1 C_{01} \,.
\end{equation}
This matrix represents the effect of going from $x=0$ to $x=1$ along $C_{01}$, making a single loop around $x=1$, and then returning to $x=0$ along the inverse of $C_{01}$. In other words, $\bar{M}_1$ is the monodromy around $x=1$ expressed in the basis of solutions normalized at $x=0$. It satisfies the same relation as $M_1$,
\[
\bar{M}_1^3 = 1,
\]
so it generates a cyclic subgroup of order three acting on the space of local solutions at the base point.

Let $P_1$, $P_2$ and $P_3$ now denote the parallel–transport operators (path-ordered products of local evolution matrices) associated with the three paths from $x=0$ to $x=2+i$ mentioned above. These operators map the boundary vector $J(0)$ to the corresponding values of $J(2+i)$ along each path. Because the paths differ only by how many times they wind around $x=1$, their transport operators are related by powers of $\bar{M}_1$:
\begin{eqnarray}
P_1 & = & P_2 \,\bar{M}_1 \,,
\nonumber \\
P_3 & = & P_1 \,\bar{M}_1 \,,
\nonumber \\
P_3 & = & P_2 \,\bar{M}_1 \bar{M}_1 =  P_2 \,\bar{M}_1^{-1} \,.
\end{eqnarray} 
Thus, to describe all three branches of the function at $x=2+i$, it suffices to compute the solution along a single path (say $P_2$), evaluate the auxiliary transport $C_{01}$ and the monodromy $M_1$, and then generate the remaining branches by acting with $\bar{M}_1$ and $\bar{M}_1^{-1}$. In this way the Frobenius-based analytic continuation, combined with monodromy, provides a systematic and efficient way to explore all possible values of a multivalued hypergeometric function at a given point.

\subsection{A bivariate example: monodromy of an Appell \(F_3\) function}\label{subsec:monodromy-2d}

We now return to the function in \eqref{eq:F3functionExample} and discuss its monodromy in more detail. In the basis
\[
J = \{1, \theta_x, \theta_y, \theta_y\theta_x\}\,F_3
\]
the differential equation takes the Pfaffian form
\[
dJ  = (M_x\, dx +M_y\, dy)\,J \,,
\]
where
\begin{equation}
M_x = \left(
\begin{array}{cccc}
 0 & \frac{1}{x} & 0 & 0 \\
 -\frac{1}{3 (x-1)} & -\frac{\frac{4 x}{3}+\frac{4}{5}}{(x-1) x} & 0 & \frac{1}{(x-1) x} \\
 0 & 0 & 0 & \frac{1}{x} \\
 0 & -\frac{y}{8 x (x y-x-y)} & -\frac{y-1}{3 (x y-x-y)} & -\frac{\frac{4 x y}{3}-\frac{4
   x}{3}+\frac{31 y}{20}}{x (x y-x-y)} \\
\end{array}
\right)
\end{equation}
and
\begin{equation}
M_y = \left(
\begin{array}{cccc}
 0 & 0 & \frac{1}{y} & 0 \\
 0 & 0 & 0 & \frac{1}{y} \\
 -\frac{1}{8 (y-1)} & 0 & -\frac{\frac{3 y}{4}+\frac{4}{5}}{(y-1) y} & \frac{1}{(y-1) y} \\
 0 & -\frac{x-1}{8 (x y-x-y)} & -\frac{x}{3 y (x y-x-y)} & -\frac{\frac{3 x y}{4}+\frac{32
   x}{15}-\frac{3 y}{4}}{y (x y-x-y)} \\
\end{array}
\right).
\end{equation}
To connect with the numerical example \eqref{eq:F3functionExample} we restrict the system to a straight line in the \((x,y)\)–space, parametrized by a complex variable \(t\),
\[
x = (2-i)\,t,\qquad y = (3+i)\,t,
\]
so that the target point \((x,y)=(2-i,3+i)\) corresponds to \(t=1\). Substituting these expressions into \(M_x\) and \(M_y\) we obtain a first–order system in the single variable \(t\),
\begin{equation}
\frac{dJ}{dt} = M_t(t)\,J(t)\,,
\end{equation}
with
\begin{equation}
M_t =\left(
\begin{array}{cccc}
 0 & \frac{1}{t} & \frac{1}{t} & 0 \\
 \frac{1+2 i}{3 i-(3+6 i) t} & -\frac{4 ((5+10 i) t+3 i)}{15 t ((1+2 i) t-i)} & 0 & \frac{1+2
   i}{(1+2 i) t-i} \\
 \frac{3+i}{8-(24+8 i) t} & 0 & \frac{16+(45+15 i) t}{20 t-(60+20 i) t^2} & \frac{3+i}{-1+(3+i)
   t} \\
 0 & \frac{3+i}{(16+8 i)-(24+8 i) t} & \frac{3+i}{(6+3 i)-(9+3 i) t} & -\frac{(375+125 i)
   t+(96+48 i)}{60 t ((3+i) t-(2+i))} \\
\end{array}
\right).
\end{equation}
We are interested in the value of \(J\) at \(t=1\). As in the univariate example, this value is not unique: there are infinitely many homotopy–inequivalent paths from the base point (which we again take near \(t=0\)) to \(t=1\), and each homotopy class defines a different branch of the multivalued solution. Four representative paths \(P_1,\dots,P_4\) are shown in the left panel of Fig.~\ref{fig:eqGraphical1}; they differ in how they wind around the singular points of the system in the \(t\)–plane.

Solving the system numerically along these four paths using the Frobenius-based continuation described in Sec.~\ref{sec:frobenius}, we obtain four distinct values of the function:
\begin{align*}
P_1:\quad &F_3\big(1,\tfrac{1}{2};\tfrac{1}{3},\tfrac{1}{4};\tfrac{1}{5};2-i,3+i\big) \approx -3.7786647 + 3.9434886\, i,\\
P_2:\quad &F_3\big(1,\tfrac{1}{2};\tfrac{1}{3},\tfrac{1}{4};\tfrac{1}{5};2-i,3+i\big) \approx -1.8122092 + 0.3425970\, i,\\
P_3:\quad &F_3\big(1,\tfrac{1}{2};\tfrac{1}{3},\tfrac{1}{4};\tfrac{1}{5};2-i,3+i\big) \approx 0.6413959 - 5.5001748\, i,\\
P_4:\quad &F_3\big(1,\tfrac{1}{2};\tfrac{1}{3},\tfrac{1}{4};\tfrac{1}{5};2-i,3+i\big) \approx 1.9764327 - 4.7372014\, i.
\end{align*}
One immediately recognizes that the value corresponding to \(P_1\) coincides with the result returned by \textit{Wolfram Mathematica}, while \(P_2\) reproduces the value obtained from \textit{Maple}.\footnote{The value corresponding to \(P_4\) can be obtained with the package \texttt{PrecisionLauricella} by choosing the option \texttt{DeltaPrescription -> I}; the choice \texttt{DeltaPrescription -> -I} gives the branch associated with \(P_1\). And the package AppellF3 \cite{Bera:2024hlq} gives the branch associated with \(P_2\).}

\begin{figure}[h]
\centering
\begin{minipage}{0.5\textwidth}
\includegraphics[width=\textwidth]{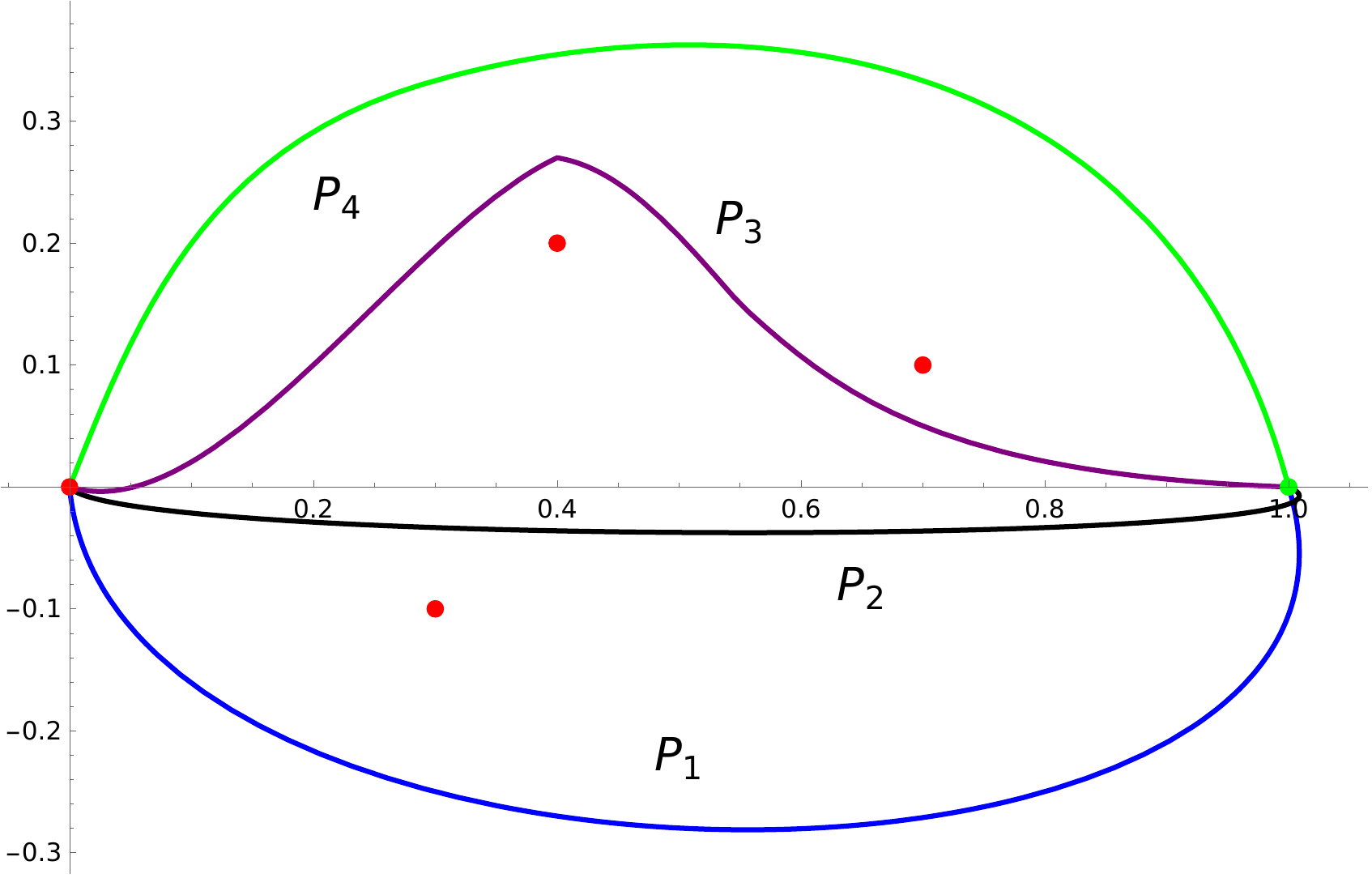}
\label{graph1}
\end{minipage}\hfill
\begin{minipage}{0.4\textwidth}
\includegraphics[width=\textwidth]{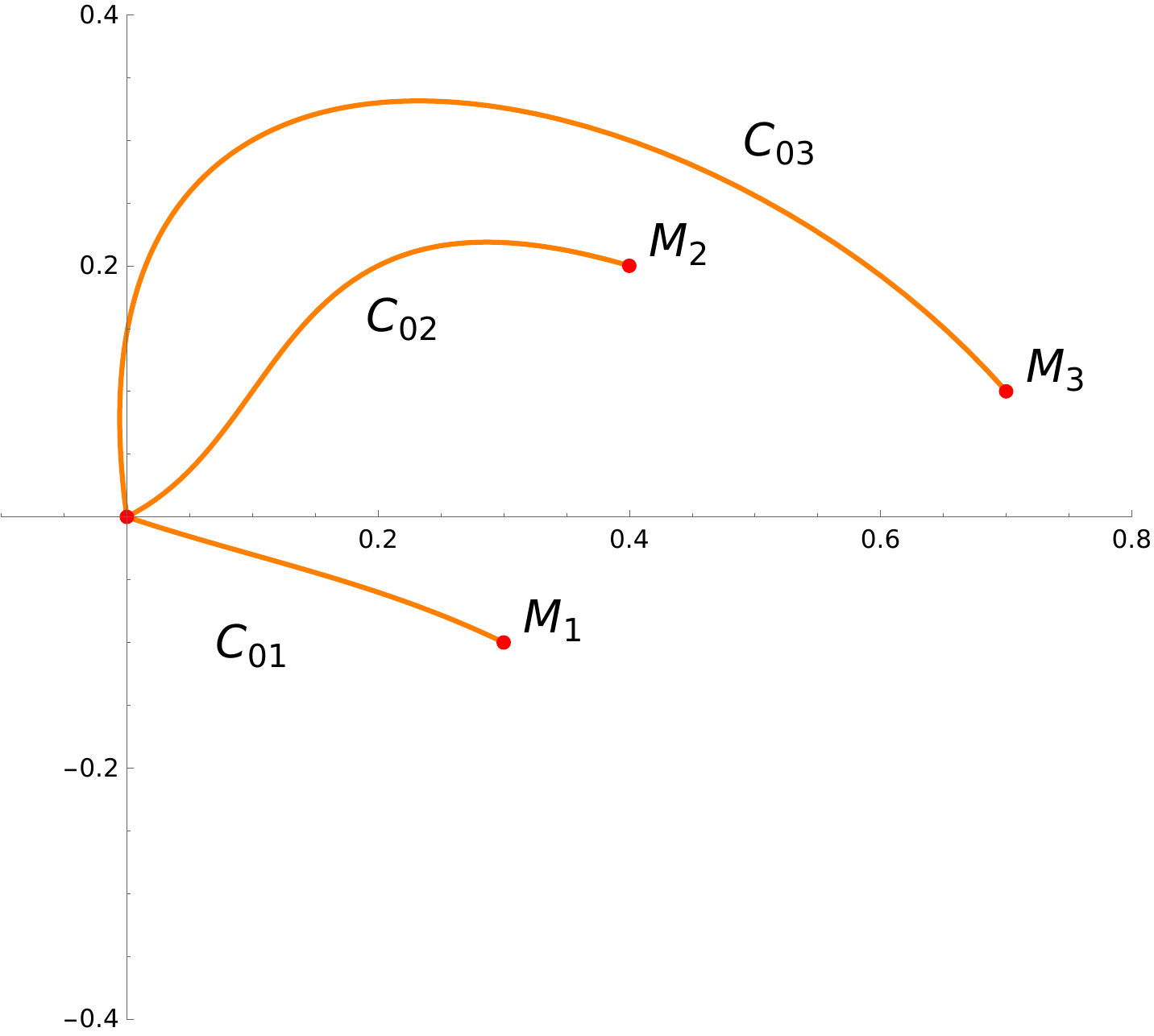}
\label{graph2}
\end{minipage}
\caption{Geometric picture of monodromy for the Appell function \(F_3\). Left: four integration paths \(P_1,\dots,P_4\) in the complex \(t\)–plane, starting from a base point near \(t=0\) and ending at \(t=1\). Red dots mark the singular points of the one–dimensional system, while the green dot indicates the target point. Right: connection paths \(C_{0i}\) from the base point to small loops around individual singularities, used to conjugate local monodromy matrices to the base point.}
\label{fig:eqGraphical1}
\end{figure}

As in the one–variable case, these different values can be related by monodromy matrices corresponding to loops around the singular points. For the restricted system in the \(t\)–plane we find three nontrivial singular points, with associated monodromy matrices
\begin{equation}
M_1 = \left(
\begin{array}{cccc}
 1 & 0 & 0 & 0 \\
 -\frac{5}{32} \left(1+(-1)^{\frac{11}{15}}\right) & -(-1)^{\frac{11}{15}} & 0 & \frac{15}{32}
   \left(1+(-1)^{\frac{11}{15}}\right) \\
 0 & 0 & 1 & 0 \\
 0 & 0 & 0 & 1 \\
\end{array}
\right),
\end{equation}
\begin{equation}
M_2 = \left(
\begin{array}{cccc}
 1 & 0 & 0 & 0 \\
 0 & 1 & 0 & 0 \\
 \frac{5}{62} \left((-1)^{\frac{9}{10}}-1\right) & 0 & (-1)^{\frac{9}{10}} & -\frac{20}{31}
   \left((-1)^{\frac{9}{10}}-1\right) \\
 0 & 0 & 0 & 1 \\
\end{array}
\right),
\end{equation}
\begin{equation}
M_3 = \left(
\begin{array}{cccc}
 1 & 0 & 0 & 0 \\
 0 & 1 & 0 & 0 \\
 0 & 0 & 1 & 0 \\
 0 & \frac{15}{346} \left((-1)^{\frac{7}{30}}-1\right) & \frac{20}{173} \left((-1)^{\frac{7}{30}}-1\right) &
   (-1)^{\frac{7}{30}} \\
\end{array}
\right),
\end{equation}
which satisfy
\[
M_1^{15} = 1, \qquad M_2^{20} = 1, \qquad M_3^{60} = 1.
\]
Each \(M_i\) describes the effect of encircling a single singular point once in the positive direction in a local basis of solutions normalized near that singularity.

To compare with paths based at \(t=0\), we introduce connection paths \(C_{0i}\), as shown schematically in the right panel of Fig.~\ref{fig:eqGraphical1}. Each \(C_{0i}\) is a path from the base point to the \(i\)-th singularity, built in the same way as any other path in our Frobenius framework (by chaining overlapping disks and local Frobenius expansions). Using these connection paths we define conjugated monodromy matrices
\[
\bar{M}_i = C_{0i}^{-1} M_i C_{0i}, \qquad i=1,2,3,
\]
which represent the action of monodromy around each singular point as seen from the base point. In terms of these matrices the parallel–transport operators for the four paths are related by
\begin{eqnarray}
P_1 & = & P_2 \,\bar{M}_1 \,,
\nonumber \\
P_2 & = & P_3 \,\bar{M}_2 \,,
\nonumber \\
P_3 & = & P_4 \,\bar{M}_3 \,.
\end{eqnarray} 

It is important to note that the connection paths \(C_{0i}\) themselves are not unique: they can be deformed within their homotopy classes, or even modified by inserting additional loops around other singularities. Such changes conjugate the matrices \(\bar{M}_i\) by products of the other \(\bar{M}_j\), but they do not alter the subgroup of the monodromy group generated by these matrices, nor the set of possible values of the function at the target point. In practice it is therefore sufficient to fix one convenient set of connection paths and work with the corresponding \(\bar{M}_i\): by combining these with the numerically computed value along a single reference path (say \(P_1\)), we can generate all other branches by explicit matrix multiplication. This example demonstrates how the Frobenius–monodromy framework turns the ambiguous notion of “branch’’ into a concrete algebraic object, and allows us to relate and systematically enumerate the different values returned by various implementations of the same hypergeometric function.

\subsection{Computation of monodromy matrices}\label{subsec:monodromy-computation}

In practice we need an efficient way to compute the monodromy matrices associated with the singular points of the one–dimensional system obtained after restricting the Pfaffian equations to a path \footnote{Sometimes, after such a reparametrization, different singularities in the original variables may collapse to the same point in the $t$–plane. For example, the singularities at $x-1=0$ and $y-1=0$ merge if we set $x = a t$, $y = a t$. However, since the matrix $M_t$ is assembled additively,
\[
\frac{dJ}{dt} = M_t J, \qquad 
M_t = \frac{\partial x_1}{\partial t} M_1
      +\frac{\partial x_2}{\partial t} M_2
      +\dots
      +\frac{\partial x_n}{\partial t} M_n,
\]
the contributions of the different singularities can still be disentangled. In this situation the corresponding singularities share the same connectivity matrix but have distinct monodromy matrices.}. In general we can write the system near a singular point $t_i$ in the form
\[
\frac{dJ}{dt} =\left( \frac{A^p_i}{(t-t_i)^{1+p}} + \frac{A^{p-1}_i}{(t-t_i)^{p}} + \dots \right)J,
\]
where $p \ge 0$ is the Poincaré rank of the singularity. When $p>0$ one may try to apply transformations that lower the Poincaré rank \cite{moser1959order,barkatou1995rational, epform2}. Under our standing assumption that all singular points of the systems we consider are regular (or can be made regular), such a transformation always exists, and we can bring the system to the form
\[
\frac{dJ}{dt} =\left( \frac{A_i}{(t-t_i)} + \dots \right)J
\]
with a single pole at $t=t_i$ and residue matrix $A_i$. In this regular case the local monodromy around $t_i$ is determined entirely by $A_i$ and its eigenvalues.

If the matrix $A_i$ is free of resonances, i.e. no two eigenvalues differ by a nonzero integer, the monodromy matrix around $t_i$ is given simply by
\[
M_i = \exp\left(  2 \pi i A_i \right).
\]
This formula follows directly from the Frobenius ansatz: in a basis where $A_i$ is diagonalizable the solutions behave like $(t-t_i)^{\lambda_j}$ near $t_i$, and a loop around $t_i$ multiplies each component by $e^{2\pi i \lambda_j}$, which is exactly the action of $\exp(2\pi i A_i)$.

In the presence of resonances (when some eigenvalues differ by integers) the situation is more subtle: logarithmic terms appear in the local solutions, and the simple exponential formula does not directly apply. Since our overall framework is numerical, one straightforward option is to introduce a small parameter shift of the eigenvalues (or of the underlying hypergeometric parameters) by a tiny $\delta$, thereby lifting all resonances; the resulting monodromy, computed via $\exp(2\pi i A_i(\delta))$, then differs from the exact one by an amount that is well within the targeted numerical precision. Alternatively, one may use purely algebraic constructions: resonances can be removed by suitable balancing transformations that adjust the spectrum of $A_i$; the monodromy can be read off from the Frobenius recurrences \eqref{eq:Frob-ansatz} by tracking how the logarithmic solutions transform under $t-t_i \mapsto (t-t_i)e^{2\pi i}$; or, finally, it can be computed directly by numerically integrating the differential equation along a small closed loop around $t_i$ and comparing the resulting fundamental matrix with its initial value. In our implementation we regard these strategies as complementary and choose between them depending on the structure of the residue matrix and the desired balance between algebraic transparency and numerical robustness.

\begin{figure}[ht]
	\centering
	\includegraphics[width=0.45\textwidth]{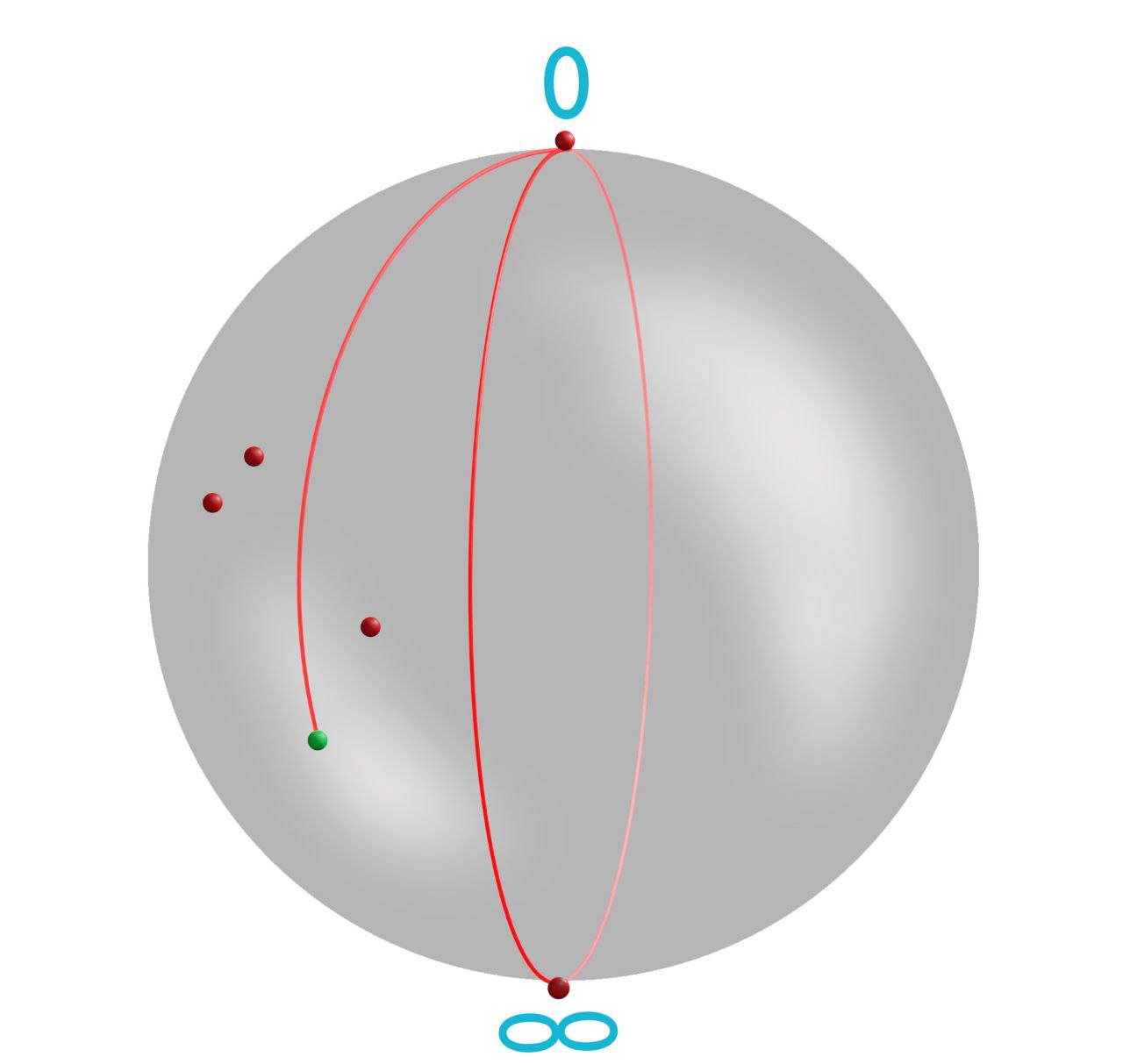}
	\caption{Riemann sphere representation of the monodromy at infinity. The north pole corresponds to $t=0$, the south pole to $t=\infty$, and the finite singularities of the differential equation are shown as red points on the sphere. The red curve represents a loop that starts at $t=0$, winds once around the singularity at infinity, and then returns to the starting point. By continuously deforming this loop on the sphere, one sees that it is homotopic to a loop that encircles all finite singularities instead of the point at infinity. Thus the monodromy at infinity can be expressed as a product of the monodromies around the finite singularities, and is not an independent element of the monodromy group.}
	\label{fig:monodromy_at_infinity}
\end{figure}

There is also a nontrivial monodromy associated with the point at infinity. It is often convenient to view the complex $t$–plane as the Riemann sphere, with $t=0$ and $t=\infty$ corresponding to the north and south poles, respectively. The singularities of the differential equation then appear as a finite set of punctures on the sphere. A closed loop encircling all finite singularities once in the positive direction is homotopic to a loop encircling the point at infinity in the negative direction, and vice versa. This situation is illustrated schematically in Fig.~\ref{fig:monodromy_at_infinity}. At the level of monodromy this implies that the monodromy at infinity can always be written as a product of the monodromies around the finite singular points (taken with suitable orientation). In other words, the monodromy representation of the fundamental group of the punctured sphere is generated by the monodromies around the finite singularities only; the monodromy at infinity is not an independent generator, but is fixed by the relation that the product of all loops on the sphere is contractible.  Passing to the complex plane with the point at infinity removed identifies loops around infinity with loops that go “around all finite singularities at once”.

A special role is played by the monodromy around the base point, which in our construction is chosen near $t=0$. Since our boundary conditions are given by the Taylor series of the hypergeometric function and its derivatives at the origin, analytic continuation along a loop that winds any number of times around $t=0$ should not change the value of the function: we are, by definition, staying on the same branch determined by this local series. In terms of our notation, let $U^{0}(t)$ be the fundamental matrix normalized at $t=0$, $b$ the boundary vector that encodes the initial conditions, $M_0$ the monodromy matrix around $t=0$, and $P$ the parallel–transport operator corresponding to a path starting at $t=0$. Then for any loop based at the origin we have
\[
P\, M_0 \left(U^{0}(0)\right)^{-1} b 
= P \left(U^{0}(0)\right)^{-1} b\,,
\]
which expresses the fact that the monodromy around $t=0$ acts trivially on the branch defined by the chosen boundary conditions. However, this ceases to be true once we first move to a different Riemann sheet by encircling another singularity with monodromy matrix $M_i$. In general $M_i$ and $M_0$ do not commute, $M_i M_0 \neq M_0 M_i$, so the combined effect of looping around $t_i$ and then around $t=0$ is different from doing it in the opposite order. In other words, while $M_0$ is invisible on the original sheet fixed by the power–series expansion at $t=0$, it becomes nontrivial on other sheets reached by analytic continuation around nonzero singularities, and its action must be taken into account when describing the full monodromy group.

\subsection{Branch cuts and canonical path construction}
\label{subsec:Brunch}
So far we have mostly ignored the question of branch cuts, but they are essential for fixing  coordinates on a chosen Riemann sheet and, consequently, for defining the principal value of a multivalued hypergeometric function.
In order to make the notion of a principal branch precise, we fix a system of branch cuts in the
space of variables and, at the same time, prescribe a canonical way of connecting the origin
with any given point $\vec{x} \in \mathbb{C}^n$. To define a hypergeometric function on multiple Riemann sheets we consider the special canonical path homotopic to a general path, such that on every given Riemann sheet it avoids cuts similar to what we did in section \ref{sec:frobenius} unless the transition to a new sheet can not be avoided. Moreover, we have a freedom in choosing the particular transition points between different Riemann sheets. The only requirement is that on each Riemann sheet our homotopic path is piece-linear path considered in \cite{Bezuglov:2025xol}.  That is, we are demanding
that for the canonical path on a given Riemann sheet the ratios
\[
\kappa_i(t) \equiv \frac{x_i(t)}{t}
\]
remain non–negative real numbers.\footnote{A detailed discussion of this choice can be found in Ref.~\cite{Bezuglov:2025xol}.}
Operationally, the package constructs a piecewise–linear path from $\vec{0}$ to $\vec{x}$ that
consists of at most four straight segments and never rotates the branch cuts of the singular
points. This ensures that, once the cuts are fixed, we can perform analytic continuation on a given Riemann sheet without crossing them.

The construction can be summarized as a four–step procedure
\begin{enumerate}
  \item \textbf{\(+\Re\)}: starting from the origin, move along the real axis to an intermediate point
        $\vec{x}^{(1)}$ obtained from $\vec{x}$ by setting all imaginary parts $\operatorname{Im} x_i$ to zero
        and replacing any negative real parts by zero. After this step all components
        $x_i^{(1)}$ are non–negative real numbers.

  \item \textbf{\(-\Re\)}: move back along the real axis from $\vec{x}^{(1)}$ to a point
        $\vec{x}^{(2)}$ by restoring exactly those negative real parts that the final point
        $\vec{x}$ should have, while keeping $\operatorname{Im} x_i = 0$. At this stage the sign
        pattern of the real parts coincides with that of $\vec{x}$, but all imaginary parts are
        still zero.

  \item \textbf{\(+\Im\)}: from $\vec{x}^{(2)}$, move parallel to the imaginary axis to
        $\vec{x}^{(3)}$ by adding all required positive imaginary parts and setting any negative
        imaginary parts to zero. Now the signs of the real parts are already correct, and the
        positive imaginary parts match those of $\vec{x}$.

  \item \textbf{\(-\Im\)}: finally, descend along the imaginary direction from $\vec{x}^{(3)}$
        to the target point $\vec{x}$, restoring the negative imaginary parts. This completes the
        path without ever changing the chosen direction of the branch cuts.
\end{enumerate}
After each projection, one of the four sign combinations $(\pm\Re,\pm\Im)$ is fixed. A fourth
segment is always sufficient to reach any quadrant, so no further steps are required. In this way
the canonical path is uniquely determined for every $\vec{x}$, and the corresponding branch of
the hypergeometric function is fixed by construction.

In the context of monodromy, this setup has two important consequences. First, monodromy
matrices are used only when we explicitly insert loops around singularities at the level of the
canonical path; this is best done in the first step. Once the path is fixed, the subsequent Frobenius continuation proceeds without
crossing any cuts, so we remain on the same Riemann sheet. Second, the connection paths
$C_{0i}$ that relate the base point to neighbourhoods of the singularities are chosen not to intersect any branch cuts. As a result, the whole scheme is self–consistent and unambiguous: the choice of cuts and canonical paths fixes the principal branch, while the monodromy matrices describe controlled transitions between different sheets. To check if two values of hypergeometric function belong to the same Riemann sheet we can for example analytically continue them avoiding cuts to the origin and check if the obtained values are the same.

\section{Confluent hypergeometric functions and irregular singularities}
\label{sec:confluent}

The framework developed in this paper assumes that all singular points of the Pfaffian system
are regular. This covers a broad class of multivariate hypergeometric functions, including all
examples discussed above. However, many important special functions arise in confluent
limits, where two or more regular singularities merge and produce an \emph{irregular} singular
point. In such cases our present method can still construct the differential system via the
Laporta-style reduction, but the Frobenius-based analytic continuation and monodromy analysis
require a genuine extension.

A standard univariate example is the Kummer (confluent hypergeometric) function
\[
M(a,b,z) = {}_1F_1(a;b;z),
\]
which satisfies
\begin{equation}
z\,y''(z) + (b - z)\,y'(z) - a\,y(z) = 0.
\end{equation}
This equation has a regular singular point at $z=0$ and an irregular singular point at
$z=\infty$ of Poincaré rank~1. For generic $(a,b)$ one finds two independent solutions with
asymptotic behaviour
\[
y_1(z) \sim z^{a-b} e^{z}, \qquad
y_2(z) \sim z^{-a}
\]
as $z \to \infty$ in suitable sectors of the complex plane. The exponential factor $e^{z}$ and
sector-dependent behaviour reflect the irregular nature of the singularity; in addition to usual
monodromy one has Stokes phenomena, governed by Stokes matrices.

On the algebraic side, the Laporta-style construction of a holonomic system for a confluent
hypergeometric function works exactly as before. Starting from the series representation of
${}_1F_1(a;b;z)$ or its multivariate confluent analogues, we can derive the relations between
derivatives, perform the reduction, and obtain a Pfaffian system. This part does not depend on
whether the singularities are regular or irregular. Likewise, we can compute numerical
solutions in regions that are bounded away from irregular points, as long as the local behaviour
is sufficiently mild.

The limitation appears when we try to treat neighbourhoods of irregular singularities themselves
or to describe the global analytic structure. The generalized Frobenius method used in
Sec.~\ref{sec:frobenius} is designed for regular singular points and constructs solutions as power
series in $(t-t_i)$ multiplied by powers and logarithms. Near an irregular singularity the natural
local form involves exponential factors and, in general, series with possibly negative powers. A
typical ansatz at $t=t_i$ would be \cite{barkatou1997algorithm, wasow2018asymptotic}
\[
U(t) = \Phi((t-t_i)^{1/s})(t-t_i)^{R}\exp\!\big(Q((t-t_i)^{-1/s})\big)\,,
\]
where $s \in \mathbb{Z}_{\ge 1}$ called the ramification index, $Q$ is a diagonal matrix containing polynomials in $(t-t_i)^{-1/s}$ without constant terms, $R$ is a constant matrix and $\Phi((t-t_i)^{1/s})$ is a formal meromorphic series in $(t-t_i)^{1/s}$.

At present our implementation does not include such irregular-singularity machinery. For
confluent hypergeometric functions we can therefore (i) construct the Pfaffian system via the
Laporta method, and (ii) obtain high-precision values in regions where the continuation path
stays away from irregular points. However, we cannot yet systematically solve the system in a
neighbourhood of an irregular singularity, nor correctly account for the associated Stokes
phenomena and “branch structure’’ of confluent functions.
In future work we plan to extend our package to cover confluent cases as well. 

\section{Conclusions}
\label{sec:conclusions}

In this work we have developed a general framework for the numerical evaluation of multivariate hypergeometric functions with arbitrary parameters and in arbitrary kinematic regions. Starting from a very broad class of hypergeometric series of several variables, we systematically derive the associated holonomic systems of partial differential equations and bring them to Pfaffian form. A key ingredient of our approach is the use of Laporta-style reduction for differential operators: by treating differential combinations as analogues of Feynman integrals and their derivatives, we construct and solve large linear systems in order to identify a minimal basis of ``master derivatives'' and express all remaining derivatives in terms of this basis.

On top of this symbolic preprocessing, we employ a modified Frobenius method to obtain high-precision numerical solutions around regular singular points and to perform analytic continuation along arbitrary paths in complex space. The continuation is organized as a sequence of overlapping disks whose radii are adapted to the local singularity structure, while a fixed prescription for branch cuts and an explicit construction of monodromy matrices allow us to control the multi-valuedness of the solutions and to access different Riemann sheets in a systematic way. Through a set of representative examples, we have shown that this framework can reproduce and connect the distinct values returned by existing computer algebra systems, and that it remains stable and efficient for genuinely multivariate functions.

Although our current implementation is restricted to functions whose differential systems have only regular singularities, the overall strategy is quite flexible. The Laporta-type construction of Pfaffian systems is already applicable to confluent hypergeometric functions, and the Frobenius–monodromy machinery can be extended once suitable local solutions around irregular singular points are incorporated. In the future we plan to generalize our methods to confluent cases, to interface them more tightly with reduction and differential-equation tools used for multi-loop Feynman integrals, and to exploit the resulting Pfaffian systems for further analytic applications, such as polylogarithmic expansions and the study of monodromy groups of higher-dimensional hypergeometric functions.

\section*{Acknowledgments}
We would like to thank V.V.~Bytev, R.N.~Lee and A.V.~Kotikov for interesting and stimulating discussions. 
The work of M.A.B. and B.A.K. was supported by the German Research Foundation DFG through Grant No.~KN 365/16-1.
The work of O.L.V. was supported by the DFG Research Unit FOR 2926 through Grant No.\ KN 365/13-2.

\appendix

\section{The \texttt{HAPC} package}
\label{appendix::HAPC}

The methods described in this paper are implemented in the \texttt{HAPC} package
(\emph{Hypergeometric Analytic Pfaffian Continuation}), written in \textit{Mathematica}.
At the present stage, \texttt{HAPC} should be regarded as a beta version and primarily as a
proof of concept: it aims to demonstrate that Laporta-style reduction and Frobenius-based
analytic continuation can be combined into a practical tool for multivariate hypergeometric
functions, rather than to provide a fully optimized production code.

The \texttt{HAPC} package can be freely downloaded from the Bitbucket repository

\url{https://BezuglovMaxim@bitbucket.org/BezuglovMaxim/hapc.git}. The entire package consists of a single file \texttt{HAPC.wl} and, provided the
path is set correctly, can be loaded into \textit{Mathematica} with
\begin{lstlisting}[language=Mathematica]
<< HAPC`
\end{lstlisting}

The central high-level routine for numerical $\ep$–expansions of multivariate hypergeometric
series is \texttt{NHypergeometrySeries}. Conceptually, it accepts the data of a hypergeometric
series (coefficients written in terms of Pochhammer symbols, summation indices, variables and
evaluation point) and returns the Laurent expansion in $\ep$ with user–specified precision.

The first argument is an association with four keys:
\begin{itemize}
  \item \texttt{"SeriesCoeficient"} – an expression giving the coefficient of
        $x_1^{\nu_1}\cdots x_n^{\nu_n}$ in terms of Pochhammer symbols
        \texttt{Pochhammer[a, m]} where the parameters \texttt{a} are linear functions of $\ep$;
  \item \texttt{"sumInd"} – the list of summation indices, e.g.\ \texttt{\{m, n, …\}};
  \item \texttt{"var"} – the corresponding list of variables, e.g.\ \texttt{\{x, y, …\}};
  \item \texttt{"rep"} – a replacement list specifying the point at which the analytic continuation
        should be evaluated.
\end{itemize}
For example:
\begin{lstlisting}[language=Mathematica]
SerData = <|
  "SeriesCoeficient" -> (Pochhammer[e, 2 m - n] Pochhammer[e, n] 
    Pochhammer[e, n])/(Pochhammer[1/2 + e, m] n! m!),
  "sumInd" -> {m, n},
  "var" -> {x, y},
  "rep" -> {x -> 2, y -> 3/2}
|>;
\end{lstlisting}

The second argument is a list $\{\ep, k\}$, where $\ep$ is the expansion parameter and
$k$ is the required order of the Laurent series in $\ep$. The last argument specifies the
requested number of decimal digits for the numerical result. For instance,
\begin{lstlisting}[language=Mathematica]
NHypergeometrySeries[SerData, {e, 3}, 20]
\end{lstlisting}
returns
\begin{lstlisting}[language=Mathematica]
(1.00000000000000000000 + 0.*10^-50 I) 
- (0.9729550745276566526 + 1.5707963267948966192 I) e 
+ (0.09043955387787492454 + 0.52391215325149202802 I) e^2
- (7.4665899801801315069 + 6.8127666264634551081 I) e^3
\end{lstlisting}
which is the $\ep$–expansion of the chosen hypergeometric series at the specified point with 20
decimal digits of precision. Additional functionality, options and worked examples are provided
in the accompanying \textit{Mathematica} notebooks distributed with the package.

Beyond the examples discussed in the main text, we performed a number of independent
consistency checks. The most stringent tests involve functions that can be expressed in terms
of multiple polylogarithms (MPLs). There exist several public packages that convert certain
classes of hypergeometric functions into MPL representations and evaluate them numerically
(see, for example, Refs. \cite{Bera:2023pyz,Bezuglov:2023owj}); some of the functions we consider lie in the overlap of their
domains of applicability. In particular, we tested Horn-type functions from the classical list,
for which MPL representations can be generated using the package~\cite{Bera:2023pyz}, as well as
four-variable Lauricella functions accessible to the package~\cite{Bezuglov:2023owj}. We additionally checked
the Lauricella–Saran function $F_S^{(3)}$, which can be evaluated numerically with the help
of package~\cite{Bera:2024hlq}. In all these cases the results of \texttt{HAPC} agree with the reference values
within the targeted precision.

We also verified various functional relations between multivariate hypergeometric functions.
As an example, we tested numerically the transformation formula
\cite{saran1955transformations}
\[
F_T(\alpha_1,\alpha_2;\beta_1,\beta_2;\beta_1+\beta_2;x,y,z) = (1-y)^{-\alpha_1}
F_1\!\left(\beta_1,\alpha_1,\alpha_2; \beta_1+\beta_2; x, \frac{z-y}{1-y}\right),
\]
where
\[
F_T(\alpha_1,\alpha_2;\beta_1,\beta_2;\gamma;x,y,z) =
\sum\limits_{m,n,p=0}^{\infty}\frac{(\alpha_1)_m (\alpha_2)_{n+p}(\beta_1)_{m+p}(\beta_2)_n }{(\gamma
   )_{m+n+p}}\frac{x^m y^n z^p}{m!n!p!}
\]
and $F_1$ is the standard Appell function. The identity holds numerically for random choices
of parameters and arguments in the domain of convergence, providing a nontrivial cross-check
of the analytic continuation and branch-handling logic. 

The correctness of a given Pfaffian system can also be checked independently by substituting the series representation of the basis functions into the differential equations and verifying that the resulting power series vanish coefficient by coefficient.

At the moment, \texttt{HAPC} does not expose a separate user-level mechanism for choosing
branches; the principal value is determined internally by the four-step path prescription
described in Sec.~\ref{subsec:Brunch}. In future versions we plan to add explicit controls for
branch selection, to refine the user interface, and to explore reimplementations of the core
algorithms in faster languages (such as \texttt{C}/\texttt{C++} or \texttt{Julia}), which should
significantly improve performance. It is also natural to incorporate more advanced variants of
Laporta reduction and to extend the current functionality to confluent hypergeometric
functions with irregular singularities, along the lines outlined in Sec.~\ref{sec:confluent}.

\bibliographystyle{hieeetr}
\bibliography{litr}

\end{document}